\documentclass[letterpaper,11pt]{article}
\pdfoutput=1
\usepackage{jheppub}
\usepackage{hyperref}
\usepackage{graphicx}
\usepackage{amsmath,color,amssymb}

\usepackage{hyperref}
\usepackage{bbm}
\usepackage{amsfonts}
\usepackage{mathrsfs}

\usepackage{amsmath,amssymb,slashed}

\usepackage{epsfig}
\usepackage{graphicx}               
\usepackage{url}
\usepackage{hyperref}
\usepackage{float}
\usepackage{pstricks}
\usepackage{color}
\usepackage{multirow}

\newcommand{\lsim}{\!\mathrel{\hbox{\rlap{\lower.55ex \hbox{$\sim$}} \kern-.34em \raise.4ex \hbox{$<$}}}}
\newcommand{\gsim}{\!\mathrel{\hbox{\rlap{\lower.55ex \hbox{$\sim$}} \kern-.34em \raise.4ex \hbox{$>$}}}}

\def\be{\begin{equation}}
\def\ee{\end{equation}}
\def\bea{\begin{eqnarray}}
\def\eea{\end{eqnarray}}
\def\bit{\begin{itemize}}
\def\eit{\end{itemize}}

\def\to{\rightarrow}

\newcommand{\beq}{\begin{eqnarray}}
\newcommand{\eeq}{\end{eqnarray}}
\newcommand{\ba}{\begin{array}}
\newcommand{\ea}{\end{array}}
\newcommand{\bal}{\begin{align}}
\newcommand{\eal}{\end{align}}
\newcommand{\bi}{\begin{itemize}}
\newcommand{\ei}{\end{itemize}}
\newcommand{\ben}{\begin{enumerate}}
\newcommand{\een}{\end{enumerate}}
\newcommand{\bc}{\begin{center}}
\newcommand{\ec}{\end{center}}
\newcommand{\bt}{\begin{table}}
\newcommand{\et}{\end{table}}
\newcommand{\btb}{\begin{tabular}}
\newcommand{\etb}{\end{tabular}}

\newcommand{\no}{\nonumber}
\newcommand{\units}[1]{\mathrm{\; #1}}

\def\lsim{\mathrel{\rlap{\lower4pt\hbox{\hskip1pt$\sim$}}
     \raise1pt\hbox{$<$}}}         
\def\gsim{\mathrel{\rlap{\lower4pt\hbox{\hskip1pt$\sim$}}
     \raise1pt\hbox{$>$}}}         

\begin{document}

\title{Detecting Superlight Dark Matter with Fermi-Degenerate Materials}
\author{Yonit Hochberg$^{1,2}$}\emailAdd{yonit.hochberg@berkeley.edu}
\author{Matt Pyle$^3$}\emailAdd{mpyle1@berkeley.edu}
\author{Yue Zhao$^4$}\emailAdd{zhaoyhep@umich.edu}
\author{Kathryn M. Zurek$^{1,2}$}\emailAdd{kzurek@berkeley.edu}
\affiliation{$^1$Theory Group, Lawrence Berkeley National Laboratory, Berkeley, CA 94709 USA}
\affiliation{$^2$Berkeley Center for Theoretical Physics, University of California, Berkeley, CA 94709 USA}
\affiliation{$^3$Physics Department, University of California, Berkeley, CA 94709 USA}
\affiliation{$^4$Michigan Center for Theoretical Physics, University of Michigan, Ann Arbor, MI 48109, USA}

\bigskip\bigskip

\abstract{
We examine in greater detail the recent proposal of using superconductors
for detecting dark matter as light as the warm dark matter limit of ${\cal O}({\rm keV})$.
Detection of such light dark matter is possible if the entire kinetic energy of the dark
matter is extracted in the scattering, and if the experiment is sensitive to
${\cal O}({\rm meV})$ energy depositions. This is the case for Fermi-degenerate
materials in which the Fermi velocity exceeds the dark matter velocity dispersion
in the Milky Way of $\sim 10^{-3}$. We focus on a concrete experimental proposal
using a superconducting target with a transition edge sensor in order to
detect the small energy deposits from the dark matter scatterings.  Considering
a wide variety of constraints, from dark matter self-interactions to the cosmic
microwave background, we show that models consistent with cosmological/astrophysical
and terrestrial constraints are observable with such detectors.  A wider range of viable models with dark matter mass below an MeV is available if dark matter or mediator properties (such as couplings or masses) differ at BBN epoch or in stellar interiors from those in superconductors.  We also show that
metal targets pay a strong in-medium suppression for kinetically mixed mediators;
this suppression is alleviated with insulating targets.
}
%
%
%
\maketitle
\flushbottom
%

\section{Introduction}

The identity of dark matter (DM) remains one of the most important
mysteries in particle physics.  In order to unlock the underlying
nature of the DM, we rely on theories to help design and
guide experiments.  The dominant theoretical paradigm of massive DM
over the last three decades has been the weakly interacting massive
particle (WIMP), and for good reason: the observed density of DM is
then naturally obtained via a freeze-out process while simultaneously
ameliorating the infamous Standard Model (SM) sensitivity to ultraviolet
physics, known as the hierarchy problem.

As a result of this focus on DM at the weak scale,
the sensitivity of experiments to such DM
has dramatically increased.  Direct and indirect detection experiments
have made impressive gains in improving constraints on DM
interaction rates with the Standard Model.

The enormous progress in the field will allow these DM experiments to push through
important benchmarks in the next five to ten years.  Ton scale
direct detection experiments such as LUX~\cite{Akerib:2013tjd}, LZ~\cite{LZ} and Xenon1T~\cite{Aprile:2012zx} already have sensitivity
to Higgs-interacting neutralino DM.  The next generations of
multi-ton experiments will have sensitivity not only to tree level
scattering but even to one loop processes, such as wino DM
scattering off nucleons through a loop of gauge bosons \cite{Hill:2013hoa}. Indirect
detection experiments are already constraining loop-generated
annihilation processes, such as thermal relic neutralino DM
annihilation into photons \cite{Cohen:2013ama,Ovanesyan:2014fwa,Bauer:2014ula}. The combination of direct and indirect
detection experiments will probe much of the viable parameter space
for the neutralino in the minimal supersymmetric Standard Model in the upcoming years.

At the same time, theoretical developments have emphasized that
compelling models of DM may be found beyond the weak scale,
especially where the dark sector is complex and displays new
dynamics. Unlike in the standard picture of DM, where the dark
matter is inert since the time that its density is set in the early
universe, such theories give rise to astrophysical and cosmological
signatures that evolve with the universe itself.  They often feature
dark sectors with multiple particles and new dark forces.  The
explosion of interest in these sectors has accompanied studies of
light hidden sectors at the LHC, `Hidden Valleys'~\cite{Strassler:2006im},
where weak scale states decay into a complex dark sector with complex
new dynamics.  Thus, even if supersymmetry is discovered at the LHC,
it in no way decreases the motivation to look for new physics beyond
the SM at a much lower scale.

Most importantly, because the masses and composition of particles in
such sectors are different than in the standard WIMP paradigm, new
experiments must be designed to search for these dark sectors. Examples are asymmetric DM from a hidden sector \cite{Kaplan:2009ag}
(natural mass scale $m_X \simeq \frac{\Omega_X}{\Omega_p} m_p
\approx 5 \mbox{ GeV}$) and mirror DM \cite{Mohapatra:2000qx,Mohapatra:2001sx}, where the masses are just below the reach of current
direct detection probes.  Going to lower masses, well-motivated theories generated radiatively from the weak
scale naturally live in the MeV-GeV mass scale~\cite{Pospelov:2007mp,Hooper:2008im,Kumar:2009bw}, as well as models
where the relic density is set via strong interactions, such as
SIMPs~\cite{Hochberg:2014dra,Hochberg:2014kqa}. Experiments are already moving toward
detecting these theories of light DM.  To this end, the detection of smaller energy deposits in the DM interaction process is required.
For example, in nuclear elastic scattering processes, the deposited energy is $E_D
\simeq {\bf q}^2/(2 m_N)$, where $m_N$ is the target mass and the momentum transfer $q \sim \mu_r v_X$
is set by the DM-nucleus reduced mass $\mu_r$ and the DM velocity
$v_X \sim 10^{-3}$.  The deposited energy on, {\it e.g.}, a
germanium nucleus is approximately 10~eV for 1~GeV DM, while past
direct detection experiments, focused on weak scale DM, were
sensitive to energy deposits between 5 and 100~keV.  In order to
access lighter DM candidates such as asymmetric DM, SuperCDMS~\cite{Agnese:2013jaa,Agnese:2014aze,Agnese:2015nto}
has lowered its energy sensitivity to 300~eV and plans to go to lower energies still.

Greater sensitivity to lighter DM for a given deposited energy can be achieved via inelastic processes,
such as electron ionization or excitation~\cite{Essig:2011nj,Graham:2012su,Essig:2015cda}.  In this case,
the process may be catalyzed if the incoming DM kinetic energy, $E_{\rm kin} = m_X v_X^2/2$, exceeds the
binding energy.  Since semi-conductors feature valence electrons with binding energies as small as a few~eV,
semi-conductor based experiments such as SuperCDMS may be used to detect MeV DM scattering with electrons.
The primary challenge is then to make heat sensors with sufficiently good energy resolution to detect 1~eV
deposits of energy on electrons.  This is a development challenge that SuperCDMS is currently taking on,
and a subject that we return to in Section~\ref{sec:detector}.
To go to even lower
scales, however, will require even lower thresholds.  Since the
threshold of a semi-conductor experiment is fundamentally limited by
the ionization energy of valence electrons, a new type of technology
will need to be developed.

In this paper we further develop the proposal laid out in
Ref.~\cite{Hochberg:2015pha}, where superconductors were considered for
accessing DM energy deposits on electrons as low as a milli-eV,
translating to sensitivity to DM with mass as low as $m_X
\sim$~keV. Cosmologically, the keV mass scale is significant because it
corresponds to the lower bound on the DM mass from the
Lyman-$\alpha$ forest~\cite{Boyarsky:2008xj} and phase space packing~\cite{Tremaine:1979we,Boyarsky:2008ju};
lighter (fermionic and thermalized) DM is inconsistent with cosmological observations (though see Ref.~\cite{Madsen:1990pe}).

There are three features which make superconductors good DM
detectors. First, ordinary metals have vanishing ionization
threshold for electrons, implying no gap and hence access (in
principle) to arbitrarily low DM energy depositions.  Second, metals
are Fermi-degenerate, meaning that the conduction electrons follow a
Fermi-Dirac distribution, having a Fermi velocity which is typically
quite substantial, $v_F \sim 10^{-2}$.  As we will see, this
Fermi-velocity is important for extracting the entire DM kinetic energy
in the scattering process.   Third, when the metal
becomes superconducting, a gap develops between the electrons in the
Fermi sea and the states into which the electrons can scatter.  The
size of this gap is small (of order a~meV), but it is crucial for
controlling the noise.  It effectively allows the decoupling of
energy deposited in vibrations in the lattice (phonons, dominated by
thermal noise) from energy deposited directly into an electron in a
hard scatter.

The energy from DM is deposited into the detector when the DM interacts with one of the electrons in the ground state of the system, namely in a Cooper pair.  When the energy deposited is larger than the Cooper pair binding energy (related to the gap), the pair is broken, and two quasiparticles are excited above the gap.  These excitations are then detected via a mechanism that we describe in detail in Section~\ref{sec:detector}.

The outline of this paper is as follows. In
Section~\ref{sec:detector} we outline the basic notion of the
detection method, and present two concrete detector designs. In
Section~\ref{sec:pauli} we describe the treatment of DM scattering
in a Fermi-degenerate medium. Section~\ref{sec:constraint} contains
various constraints on DM scattering with electrons. In
Section~\ref{sec:models} we discuss several particular models:
Scalar and vector mediation is considered in
Section~\ref{ssec:scalar}; a kinetically mixed hidden photon is
considered in Section~\ref{ssec:hiddenU1}, including in-medium
effects; dipole interactions of DM are tackled in
Section~\ref{ssec:dipole}; and milli-charged DM is discussed in
Section~\ref{ssec:mili}. We conclude in Section~\ref{sec:conc}.

\section{Detection principle and design}\label{sec:detector}

We begin by presenting the underlying idea behind our proposed
detection method. After establishing the basic notion for detecting
${\cal O}$(meV) energy depositions, we present concrete
experimental detector designs that could be sensitive to this energy range.

\subsection{Detection principle}\label{ssec:principle}

When searching for DM with mass heavier than $\sim 10$  GeV with elastic scattering, nuclear targets have three main advantages:
First, DM elastic scattering 
has a rate that scales as the reduced mass of the DM-target system,
$\mu_r$, which suppresses the scattering rate on electrons compared
to that on nucleons. Second, the maximum deposited energy in an
elastic recoil off of a target at rest is
\begin{equation}
E_{D} = \frac{\mu_{r}^2}{2m_{T}}v_{X}^{2}\,,
\label{eq:Ed_v0target}
\end{equation}
which is maximized when the target has mass equal to that of the DM
$X$.  Here, $m_T$ is the target mass and $v_X \sim 10^{-3}$ the DM
velocity.  Thus, 100~GeV DM produces nuclear recoils of ${\cal
O}(10\;{\rm keV})$, but e$^{-}$ recoils of only ${\cal O}({\rm
eV})$. Third, backgrounds in direct detection experiments, such as
Compton scattering, feature mainly an electron ionization component;
thus, discriminating nuclear recoils from electromagnetic activity
acts as a major discriminant for reducing backgrounds.

As the DM mass drops below the mass of the nuclear target, around
10's of GeV, Eq.~\eqref{eq:Ed_v0target} indicates that the deposited
energy is suppressed by $m_X^2/m_T^2$ compared to the case of DM
heavier than the target.  For example, sensitivity to 1~eV nuclear
recoils allows reach to 100~MeV DM.  Searching for such 1~eV nuclear
recoils from 100~MeV DM scatterings has motivated SuperCDMS's push
to lower thresholds~~\cite{Agnese:2013jaa,Agnese:2014aze,Agnese:2015nto}.  In
addition, utilizing a lighter nuclear target, such as $^4$He, is
also advantageous in searching for lighter DM~\cite{mckinsey}.

To access even lighter DM, electron targets are preferred.  In this case, an energy deposition  sensitivity of 1~eV
corresponds to probing DM models with mass down to roughly 1~MeV.  Of course,
this sensitivity can only be achieved if the energy deposit exceeds
the binding energy of the electron.  In a xenon atom, the binding
energy of the outermost electron is 12~eV, while in germanium the
band gap is 0.7~eV. Thus, the binding energy of
electrons in atomic targets and semi-conductors fundamentally limits
access to DM candidates with mass below an MeV.  To access such
candidates, we need a material with a gap smaller than $m_X
v_X^2/2$, which corresponds to ${\cal O}({\rm meV})$ energy for DM
at the warm DM limit of ${\cal O}({\rm keV})$.  Metals (including
metals in a superconducting phase) and superfluids are examples of
materials that feature a small or no gap, and as such can be
appropriate.

Thus materials with a ${\cal O}({\rm meV})$ gap and sensitivity to ${\cal O}({\rm meV})$ energy depositions may allow for the detection of DM at the ${\cal O}({\rm keV})$ mass scale. However, even when it is energetically possible for DM to catalyze a reaction---when the DM kinetic energy exceeds the gap of the material---kinematics may still forbid the scattering.
To see this, consider the deposited energy on a target (electron or nucleus) in terms of the momentum transfer of the process ${\bf q}$:
\beq
E_D \simeq
\frac{1}{2}\left(\frac{{\bf q}^2}{m_T} + 2 {\bf q} \cdot {\vec v}_{i,T}
\right) + \delta \,, \label{FullED}
\eeq
where $v_{i,T}$ is the initial velocity of the target and $\delta$ (defined to have a positive sign for bound electrons) is the gap of the system.
The first term in Eq.~\eqref{FullED} is the usual energy deposition for elastic scattering on targets at rest, just like Eq.~\eqref{eq:Ed_v0target}.
The third term takes into account that fact that DM may catalyze an inelastic process, releasing the binding energy of the target electron.
The second term is the one we wish to focus on: it is the effect of the target's initial velocity on the total amount of DM kinetic energy
that can be absorbed.  Even if the first term is small ({\it e.g.} $\sim \mu$eV for a keV DM scattering on an electron), the second term may allow
extraction of the entire kinetic energy of the DM. 
In the metal and superfluid targets we are most interested in here, the velocity
of the target is a property of the target ground state, and is due to Fermi statistics.  (We note that semiconductor and atomic targets feature an electron velocity of similar size.)

To illustrate this point, consider two relevant limits, when the DM
is heavier or lighter than the target. (For the purpose of this
illustration, we neglect the band gap, assuming it is substantially
smaller than the DM kinetic energy.) When the DM is much heavier
than the target, the center of mass frame is approximately the DM rest
frame. In this frame, the collision between the DM and the
target barely changes the DM velocity, and the target initial and
final state velocities have the same magnitude but are in opposite
directions. In the lab frame, the target velocity changes at most
from $v_{i,T}$ to $(v_{i,T}+2v_{X})$, with opposite direction. Thus
the maximum energy deposition can be written as $E_{D}^{\rm
max}=\frac{1}{2}m_T[(v_{i,T}+2v_{X})^2-v_{i,T}^2]$, and the momentum
transfer is $2 m_T v_X$.  When $v_{i,T} \gg v_X$, this reduces to
$E_D^{\rm max} \simeq 2 m_T v_{i,T} v_X$.  On the other hand, in the
limit that the DM is much lighter than the target, the maximum
energy deposition is obtained when the DM is fully stopped by the
target.  For example, a target with velocity
$(0,\sqrt{v_{i,T}^2-v_X^2/4},v_X/2)$ can fully stop a DM particle
with velocity $(0,0,v_X)$, and the momentum transfer in this case is
simply the DM initial momentum, $m_X v_X$.  Since the deposited
energy is approximately $\frac{1}{2} m_X v_X^2$, the experiment must
have meV energy resolution in order to be sensitive to keV mass DM.

What is the typical target
velocity in the (nearly) gapless materials we consider?  In a metal like aluminum, the valence
electrons have a Fermi momentum $p_F = 3 \mbox{ keV}$, giving rise to a Fermi
velocity for electrons of $v_F \simeq 10^{-2}$.   Note that this effect is purely due to Pauli blocking and Fermi statistics in a degenerate medium at low temperature. Superfluids such as Helium-3, where the nucleus has half-integer spin, also display
Fermi degeneracy.  In the case of Helium-3, the Fermi energy is $E_F
\simeq 4 \times 10^{-4} \mbox{ eV}$, giving rise to a Fermi velocity of
the Helium-3 nucleus of order $v_F \simeq 10^{-6}$.  For a typical
momentum transfer of $m_X v_X$, the second term in Eq.~\eqref{FullED}
never dominates in Helium-3, however, and instead the scattering
proceeds via the ordinary nuclear recoil process.

Note that the electron velocity does play a role in DM scattering off electrons in a semi-conductor or noble gas as well~\cite{Essig:2015cda}. In such scatterings, for DM heavier than an MeV, the electron velocity is not a necessary ingredient to catalyze the process and extract all of the DM energy in the scattering, but nevertheless it does impact the kinematics, since the electron velocity, $v_{i,T} \sim \alpha\; Z$ with $Z$ the electric charge of the nucleus, is larger than the velocity of the DM.

An alternative route towards detecting DM as light as a keV is to take advantage of the gap,
with inelasticity catalyzing the scattering.  This is evident from Eq.~\eqref{FullED}: even
if the first two terms are below a meV, as long as the kinetic energy of the DM exceeds the
gap, the DM energy may be absorbed by exciting an electron above the gap.  For the metal
targets we are interested in for the rest of this paper, when the metal enters the
superconducting phase, a sub-meV gap appears.  Since this gap is below detectable
energies for the devices we consider, we ignore the presence of the gap, and focus
on elastic scattering.  Also note that while the presence of the superconducting gap
is not important for the scattering process itself, its existence means that athermal
phonons and quasiparticles have very long lifetimes, and as such can potentially be
collected before they thermalize. Thus in the systems we consider, detection of DM operates via
the breaking of Cooper pairs in a superconducting target. We consider this idea in more detail next.

\subsection{Detector design with milli-eV sensitivity}\label{ssec:design}

\begin{figure}[t!]
\begin{center}
\includegraphics[clip,width=0.495\textwidth]{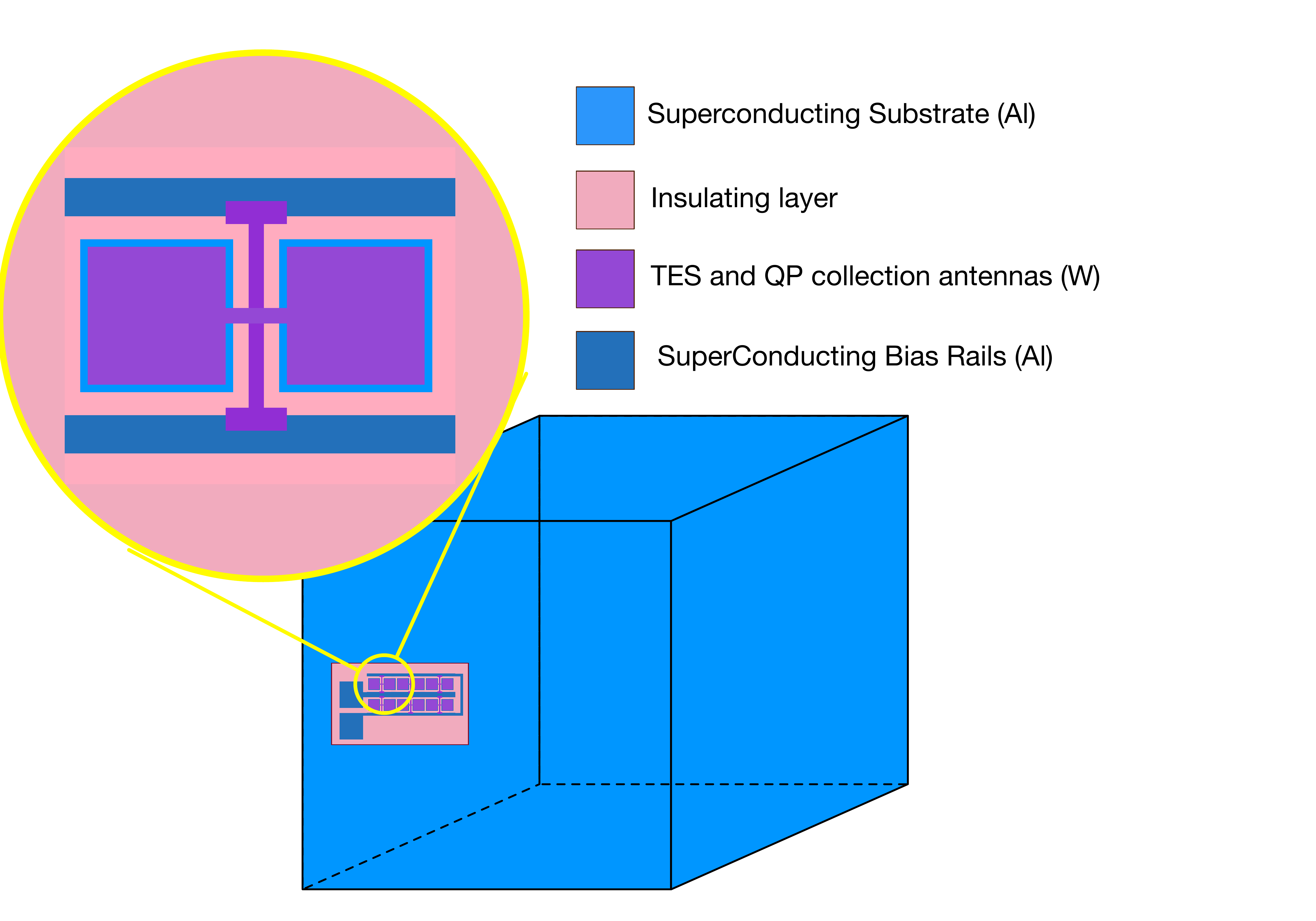}
\includegraphics[clip,width=0.495\textwidth]{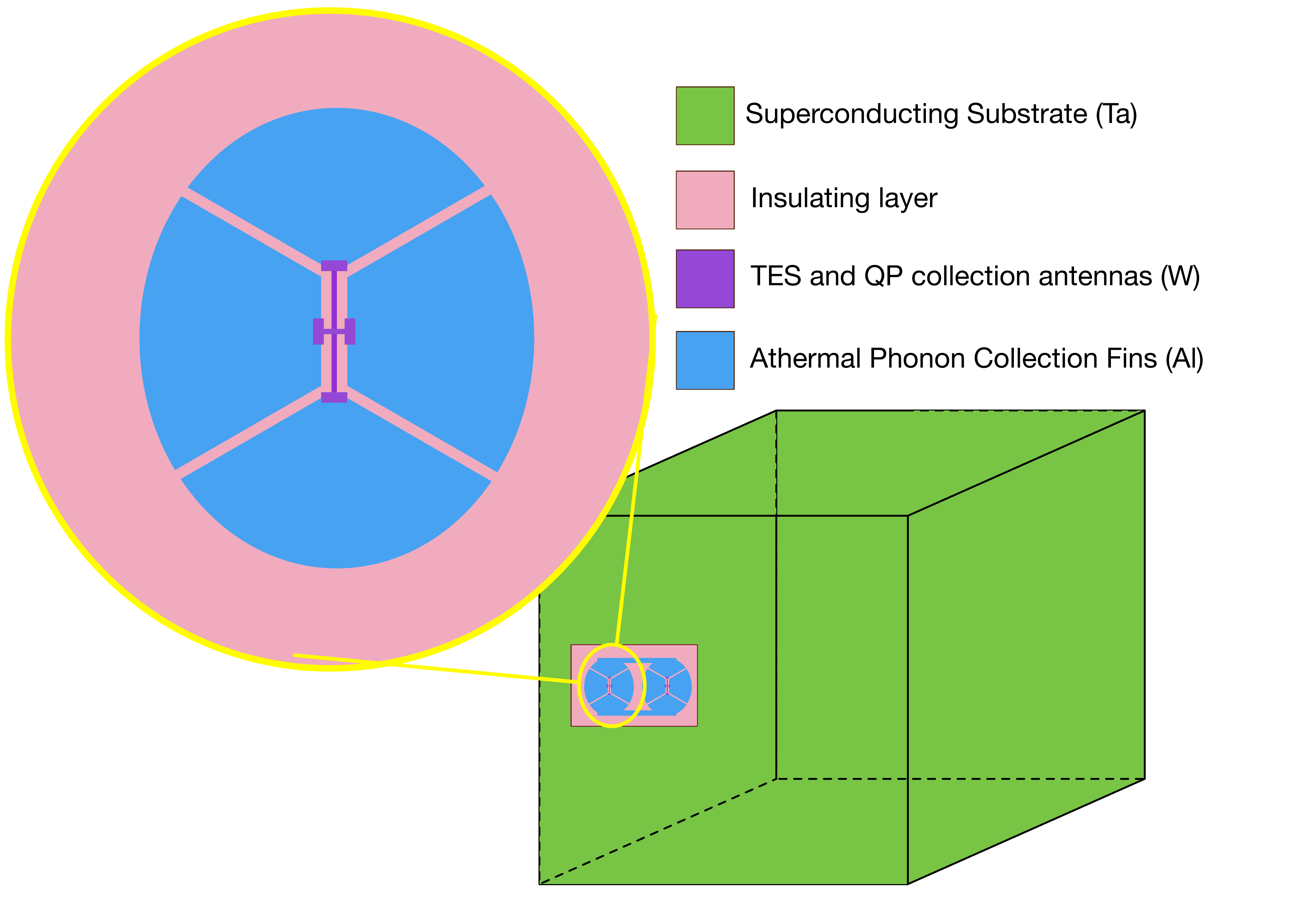}
\caption{ Schematic designs for superconducting detectors that are sensitive to  DM-electron scattering. {\bf Left:} Quasiparticles produced by a recoiling e$^{-}$ in a large aluminum arbsorber are collected by tungsten quasiparticle collection fins and then their energy is sensed by a TES.
{\bf Right:} Athermal phonons produced by a recoil e$^{-}$ in a large tantalum absorber are collected by aluminum collection fins and then their energy is sensed by a TES.}
\label{fig:schem}
\end{center}
\end{figure}

Our detector concept is based on collecting and concentrating long lived athermal excitations from DM
interactions in a superconducting target absorber onto a small volume (and thus highly sensitive) sensor.
The collection and concentration of long lived excitations is a general concept that has been a core
principle of detector physics, from ionization in semiconductor CCDs to athermal phonon collection in CDMS.
Here we propose that this general detection philosophy be applied
in large volume (very pure, single crystal) superconductors to search for DM with mass
as low as the warm DM limit of a keV using standard superconducting sensor technology
that has been pushed to its ultimate theoretical sensitivity.  A schematic of two proposed
detector concepts for light dark matter, that we describe in greater detail through the
remainder of this section, is shown in Fig.~\ref{fig:schem}.

Detection of dark matter in such detectors is comprised of a three part process:
\begin{itemize}
  \item {\it Dark Matter Scattering on Target Absorber and Subsequent Excitation Production.}  A DM particle scatters off an e$^{-}$ in the target metal or superconducting absorber.  In subsequent interactions, the recoil energy is converted into long lived athermal phonons and quasiparticles.
\item {\it Collection of Excitations.} The resulting excitations must be collected and concentrated onto a small volume (and thus very sensitive) sensor; this is typically done via `collection fins' on the surface of the absorber that efficiently collect the energy from the excitations in the absorber.
\item {\it Measurement of Excitation Energy/Quanta.}  After collecting the excitations, they must be measured.  In the devices we consider, this is typically done via a transition edge sensor (TES) or a microwave kinetic inductance device (MKID).
\end{itemize}

In the remainder of this section we describe in detail 2 potentially feasible detector designs; each of the following three subsections is dedicated to
{\em absorption and excitation production}, {\em collection} and {\em measurement}.  The theorist interested primarily in a calculation of the reach of these
detectors, as well as theories of DM that could be detected with such devices, can move directly to Section~\ref{sec:pauli}.

\subsubsection{Excitation production in superconductors}

Superconductors are excellent candidates for detection of light dark matter because the
absence of any unoccupied electronic states within the superconducting band gap $\Delta$ of the
Fermi surface means that both quasiparticles near the gap edge and athermal phonons with energy
below the gap are long lived, and thus potentially amenable to measurement.

The valence (conducting) electrons at low temperature in a
metal are well-described by a Fermi-degenerate distribution.  In a
metal like aluminum, the Fermi energy is 11.7~eV, and the
corresponding number density of conducting electrons is
\begin{equation}
n_e = \frac{(2E_F m_e)^{3/2}}{3\pi^2}\,.
\end{equation}
This is the reservoir of electrons that are available to become
superconducting once the temperature of the metal drops below the
critical temperature $T_c$ and the electrons enter the
superconducting phase; it is also these electrons that are available
for DM scattering in the target metal.

The superconducting phase is entered when it is energetically
favorable for near Fermi energy electrons to bind into pairs, known as
Cooper pairs. This ground state of the system is then highly correlated,
and free electrons are no longer the correct degrees of freedom of
the system. (For a review of superconductivity, see {\it e.g.}
Ref.~\cite{Tinkham}.)  The binding energy $2\Delta$ of these pairs is
typically quite small, {\it e.g.} of order 0.6~meV in aluminum, and
their correlation length $\xi_0=  v_F/(\pi \Delta)$
macroscopic, {\it e.g.} of order a micron in aluminum.
A DM particle interacts with these valence electrons bound into
Cooper pairs in the superconducting target. For the purpose of the
rate calculation carried out in Section~\ref{sec:pauli}, the important
point is that as long as the energy deposited in the DM-electron
scattering well-exceeds the binding energy of the Cooper pair, the
DM-electron scattering rate can be approximated via energy deposit
onto free electrons in a Fermi degenerate free-electron
sea~\cite{Tinkham}. As the energy deposited approaches the Cooper
pair binding energy, a `coherence factor', analogous to a form
factor, takes into account the impact of the coherence from the
Cooper pairing phenomenon; we discuss this further in the next section.

The initial interaction between the Cooper pair and the DM creates
two quasiparticle excitations from the ground state. As discussed in
Ref.~\cite{Kurakado_QPcascades}, for  $E \lsim 100\; \rm{ meV}$,
thermalization of these excited quasiparticles occurs predominantly
via athermal phonon production. Since the phonon phase space scales
as $E^2$, the produced phonon distributions tend to be hard; on
average most of the excess energy of the quasiparticle is converted
to a single athermal phonon quanta.  Then, as long as the produced
athermal phonon has $E > 2\Delta$, it will break an additional
Cooper pair, and so on. At the end of the cascade process, the total
kinetic energy of the recoil has been converted into $\sim 60 \%$
quasiparticle potential and kinetic energy, with the remaining
energy in athermal phonons with a distribution that is strongly near
the 2$\Delta$ cutoff in the superconductor~\cite{Pyle:2012hma}.

Once created, what are the dynamics of these excitations? Is it possible to collect them before they thermalize?  We address this important issue for quasiparticles and athermal phonons next.
\begin{enumerate}
\item {\bf Quasiparticle Dynamics.}
Extremely pure single crystal aluminum is very unique in that electronic excitation scattering lengths of $\sim$1.5~mm have been measured at temperatures of 4K (with residual resistance ratio ${\rm RRR}\equiv R(300{\rm K})/R(4.2{\rm K})\gsim 10^{5}$)~\cite{AlScattering}. Furthermore, these measured scattering lengths should underestimate the scattering length of aluminum quasiparticles in a crystal of similar quality at dilution fridge temperatures of $\sim 6$mK for two reasons:
\begin{itemize}
    \item The thermal phonon population is significantly smaller at temperatures of 6mK compared to 4K, and thus phonon up-scattering rates, which can be non-negligible at 4K in extremely pure aluminum crystals, are completely suppressed.
    \item Quasiparticle scattering rates off of impurities are suppressed compared to those of normal electrons.
\end{itemize}
Thus, for a 5mm aluminum crystal, quasiparticle propagation is essentially ballistic.

The lifetime of quasiparticles in very high quality single crystal
aluminum has not been measured to our knowledge.  A 2~ms quasiparticle lifetime has been measured in 100~nm thick
aluminum MKIDs~\cite{0811.1961}.  However, this is probably too conservative since the RRR of the aluminum films used was only
3.3 and the MKID lifetime was found to have a strong dependence on dislocation density. Furthermore, the surface to volume ratio in these films is much larger
than those found in the 5~mm cubic absorbers that we are considering and thus any recombination/trapping on the surfaces is enhanced in this thin film device
compared to what we would expect. Thus, for purpose of our calculations, we'll assume a 20ms lifetime for quasi-particles in single crystal aluminum. Together with the quasiparticle
group velocity~\cite{Ullom},
\begin{equation}
    v_{\rm QP} = v_{F}\sqrt{1-{\left(\frac{\Delta}{\Delta+k_{B}T}\right)}^{2}} \sim 10^{-3}-10^{-2}
\end{equation}
(depending on the quasiparticle temperature $T$), this suggests that for a 5~mm aluminum absorber, quasiparticles would bounce off the single
crystal surface $> 10^{6}$ times before recombination.

\item {\bf Athermal Phonon Dynamics.} Due to the lack of electronic states within the superconducting bandgap, elastic and inelastic scattering of athermal phonons with energy below 2$\Delta$ through electron-phonon interaction is impossible to lowest order in the absence of quasiparticles. Consequently, athermal phonon dynamics in high quality single crystal superconductors at $T\ll T_{c}$ mirror those found in detector grade semiconductor crystals.

In particular, elastic scattering will be driven by isotopic and impurity scattering~\cite{Tamura_IsotopeScattering}. Since natural aluminum (Al) and tantalum (Ta) are almost entirely composed of a single isotope, elastic isotopic scattering should be negligible in these superconductors. Simultaneously, impurity scattering in these materials is minimal due to the use of very pure single crystals (float zone refining is easily implemented in single crystal metals~\cite{AlScattering}). Thus---as was the case for excited quasiparticles in aluminum---athermal phonons in aluminum and tantalum should be ballistic for $\cal O (\rm{1\; cm})$-sized absorbers.  We note that this conclusion is seemingly in conflict with athermal phonon propagation studies done in single crystal lead~\cite{Wolfe_PbScattering1, Wolfe_PbScattering2}, however this is quite expected.\footnote{
The reasons are as follows: (1) The source of athermal phonons for their propagation measurements was a 30-70K thermal hot spot in a copper film on the surface of the lead crystal. Consequently, there was also a  large non-equilibrium source of quasiparticles that was shown to completely dominate scattering~\cite{Wolfe_PbScattering2}. (2)
 The athermal phonon scattering measurements were done at  $3<T_{c}/T<5$, and thus there was a non-negligible fraction of equilibrium quasiparticle scattering that would not be present at $T_{c}/T \sim 100$.
(3) Pb naturally has large isotopic scattering not present in aluminum. (4) Isotopic and impurity phonon scattering rates scale as the inverse Debye temperature cubed~\cite{Tamura_IsotopeScattering}. 
As a result, even for similar impurity levels the phonon scattering rate would be suppressed by two orders of magnitude in aluminum compared to lead.
}

In the bulk of these crystals, the lifetime of the athermal phonon will be limited by phonon anharmonic decay~\cite{Tamura_DownConversion}, in which a phonon splits to two phonons. Since the 3$^{\rm rd}$ order elastic constants (effectively, the coupling constant for this phonon splitting process) in aluminum are similar to those found in germanium and silicon, we expect the anharmonic decay lifetime for a 4K phonon to be of $\cal O (\rm{1 \;s})$. (The equivalent numbers for tantalum could not be found in the literature, but we assume they are of similar size.) We thus estimate more than $2 \times 10^{5}$ surface bounces for athermal phonons in a 5~mm absorber.  Of course, with such a large number of potential surface bounces, phonon down-conversion at the surface is most likely the dominant thermalization process. Such processes depend critically on the exact surface preparation and are thus difficult to estimate. For example, SuperCDMS has some evidence that a 30~nm amorphous silicon layer on a germanium crystal down-converts athermal phonons every $\sim$~250 bounces but on bare germanium surfaces there is only a lower limit, of 1250 bounces~\cite{SuperCDMS_unpublished}. For the purpose of estimating athermal phonon collecting detector sensitivity in both aluminum and tantalum absborbers, we will use this bare germanium lower limit.
\end{enumerate}

In summary, the quasiparticle excitations in very pure single crystal aluminum and the athermal phonon excitations in aluminum and tantalum are very likely to have characteristic lifetimes $\gsim 2$~ms and scattering lengths potentially much larger than a mm.
As a result, they have excellent potential as target material for excitation-sensitive detector technology.

The scattering length for both quasiparticles and athermal phonon bounds the size of the absorber, which we consequently take to be of order $\sim(5\; {\rm mm})^3$.
A large number of these small detectors would then be placed in parallel in order to obtain large exposure.

Having established the longevity of excitations in the absorber, we move to discuss collection and concentration of these excitations.

\subsubsection{Collection and concentration of long lived excitations}

The second step of these long-lived excitation detectors
is to collect and concentrate the excitations from the absorber into the much smaller sensing region. In semiconductors like germanium and silicon, this is largely trivial; since the electron-hole pair are electrically charged, they can be drifted towards the sensor region by biasing the sensing region at an appropriate voltage.  Such techniques are unfortunately impossible within a superconductor due to the perfect shielding of both electric and magnetic fields. (Furthermore, athermal phonons are electrically neutral.)

We follow the spirit of design concepts first laid out in superconducting tunnel junctions (STJs)~\cite{Booth_QPcollection_STJ}
and later by CDMS using athermal phonons~\cite{Irwin_QPcollectionDetector}. We simply allow the long lived excitations to randomly
propagate within the absorber, and instrument a small fraction of the overall surface of the absorber with a material that has a
high probability, $f_{\rm trap}$, of collecting/trapping the excitation upon contact.

When propagation is ballistic, an excitation will, on average, be
collected after $\frac{A_{\rm absorber}}{A_{\rm collect}}
\frac{1}{f_{\rm trap}}$ bounces, corresponding in a cubic absorber
to a collection time of~\cite{sabine}
\begin{equation}
    \tau_{\rm collect} =  \frac{4 V_{\rm absorber}}{\langle|v|\rangle A_{\rm collect}}\frac{1}{f_{\rm trap}}\,,
\end{equation}
where $A_{\rm absorber}$, $V_{\rm absorber}$, and $A_{\rm collect}$  are the total absorber surface area, the total absorber volume, and the instrumented collection area respectively, with $\langle|v|\rangle$ the average excitation velocity.

The excitation collection process competes with annihilation processes (either phonon anharmonic decay or quasiparticle recombination), and thus the average excitation collection efficiency is given by
\begin{equation}
    f_{\rm collect} = \frac{\tau_{\rm life}}{\tau_{\rm life}+\tau_{\rm collect}}\,.
\end{equation}
The benefit of very large excitation lifetimes ($\tau_{\rm life}$) is now clear. It allows
one to achieve high excitation collection efficiencies even for very large ratios of
$V_{\rm absorber}/A_{\rm collect}$.  Since sensor sensitivity scales with the size of
the sensor and thus with $A_{\rm collect}$, large excitation lifetimes allow one to
simultaneously satisfy the requirements of large exposures and very low energy thresholds
in optimized devices.

When excitation propagation is diffusive, collection probability and collection times will
be highly dependent on the location of the generated excitation. In particular, excitations
generated far from an instrumented surface will have long collection times and potentially
very poor collection probabilities. To minimize these issues, as we already commented above, we require that the absorber
be of such a size that the excitations are ballistic; thus the absorber must be relatively small (a cm or smaller).

For quasiparticle collection, the standard collector material is either a superconductor
with transition temperature $T_{c,{\rm collect}}$ that is less than that of the absorber,
$T_{c,{\rm absorber}}$, or simply a normal metal. In both cases, the collector material
is placed in direct contact with the absorber, producing a proximitized region where the
critical temperature 
(and thus the gap) is between that of the absorber and the
collector. Within this suppressed bandgap region, a portion of the quasiparticle's original
potential energy will be converted into kinetic energy, and the likelihood of inelastic phonon
production will be significantly enhanced. If this occurs, the quasiparticle will then be
trapped within the collection volume.

The thickness of the collection film $\ell_{\rm collect}$ is a result of optimizing
two competing effects. On the one hand, one would like large $f_{\rm trap}$. Since
the phonon production rate scales as the excess quasiparticle energy cubed~\cite{Kaplan_QPScattering}, and the
bandgap suppression scales roughly linearly with thickness~\cite{Blamire_ProximityTc},
 the phonon dropping rate scales as $ \ell_{\rm collect}^{\sim 3}$. Furthermore, since
 the amount of time spent in the film scales as $\ell_{\rm collect}$, we have
 $f_{\rm trap} \propto \ell_{\rm collect}^{\sim 4}$.  (Such large power law scalings
 of $f_{\rm trap}$ with $\ell_{\rm collect}$  are consistent with previous
 experiments~\cite{Moore_AlTi_StripDetectors}). On the other hand, we would like
 to maximize the efficiency of energy collection and thus minimize the energy of
 the phonons that are released in the trapping process, which suggests thinner films. In the quasi-particle collection interface between 35~nm tungsten and 350~nm aluminum films in SuperCDMS devices, $f_{\rm trap}$ is within the range of $10^{-3}$-$10^{-4}$~\cite{Yen_QPdiffusion}.  This suggests that $f_{\rm trap} \sim 0.1$ could be achievable with 100-200~nm thick collection films, which leads to an estimate of 87\% for $f_{\rm collect}$, if one places twelve 225~$\mu {\rm m}^{2}$ size quasiparticle collection fins on the surface of the aluminum absorber. We estimate that the average quasiparticle potential energy remaining after sub-gap phonon emission for trapping, $f_{\rm E\,Remain}$, will be greater than $90\%$. 

For athermal phonon collection, we again use a superconducting collecting film
with $T_{c,{\rm collect}} < T_{c,{\rm absorber}}$, but which is now electrically
isolated from the large-volume superconducting absorber by an insulating layer
(SiNx, SiOx and Al$_2$O$_3$ are all viable options). With this configuration,
athermal phonons with energies $2 \Delta_{\rm collect} < E < 2 \Delta_{\rm absorber}$
will ballistically travel throughout the absorber, but will annihilate within the
collector film, storing their energy in quasiparticles. To collect a large fraction
of the athermal phonons, the difference in the $T_{c}$ between the two materials
should be significant. Some viable possibilities are tantalum(absorber)/aluminum(collector) and aluminum(absorber)/titanium(collector).
This greater freedom in the absorber/collector duos for athermal phonon collecting detectors
is due to the fact that the requirements on the absorber in this case are much less constraining since phonon scattering lengths and lifetimes should be excellent for the vast
majority of single-element superconductors.

After the athermal phonons are converted into quasiparticles within the aluminum collection fin, they diffuse until they are absorbed by the connected TES~--- just as in CDMS athermal phonon detectors. This diffusion processes introduces an additional energy loss mechanism due to quasiparticle trapping, that has been well studied for SuperCDMS detector geometries~\cite{Yen_QPdiffusion}. We estimate the quasiparticle collection efficiency to be $\sim0.65$ for the aluminum collection fin geometry used in the proposed athermal phonon detector (see Table~\ref{tab:APQPDevice}).

\subsubsection{Measurement}
\label{sec:TES}

With the athermal excitations from the target now concentrated, all that is necessary is to read them out with a sensor of the appropriate sensitivity; in essence, one can just re-purpose a single-infared-photon-sensitive detector. Unfortunately, the required ${\cal O}(1\;{\rm meV})$ sensitivities have not yet been achieved experimentally with any technology. For both Transition Edge Sensors (TES) and Microwave Kinetic Inductance Devices (MKID) though, such sensitivities are theoretically possible. Furthermore, in both cases, engineering solutions (though extremely challenging) have been proposed, which could allow the theoretical sensitivities to be realized.  Below, we detail the engineering challenges for the TES.

The TES is a superconducting film that has been artificially stabilized through electro-thermal feedback at an operating point just within the superconducting transition.  Biased in this manner, very small changes in the temperature of its electronic system can produce substantial changes in resistivity, which are then measured~\cite{Irwin_TES}.  The theoretical energy resolution (squared) of the TES is given by
\begin{eqnarray}
    \sigma_{\rm E}^{2}  = \int_{0}^{\infty} d\nu \frac{4}{S_{\rm{p,tot}}(\nu)}\ \sim\ 4 k_{B} T_{c,{\rm TES}}^{2} C \frac{\sqrt{n_{\rm Tb}/2}}{\alpha}\,,
\label{eq:resTES}
\end{eqnarray}
where $S_{\rm{p,tot}}(\nu)$ is the total noise referenced to TES input power, $T_{c,{\rm TES}}$ is the transition temperature of the biased system and $C$ is the heat capacity of the TES.  Here $\alpha$ is the unitless measure of sensor sensitivity defined as $ \frac{T}{R} \frac{\partial R}{\partial T}$ at the TES operating point, and can take values in the range of 20-200, depending on the TES film.
Finally, $n_{\rm Tb}$ is the temperature scaling exponent on the thermal power which flows between the TES and the heat bath.
To gain intuition into Eq.~\eqref{eq:resTES}, note that the energy variance for a heat capacitor $C$ coupled to a thermal bath via conductance $G$ is $4 k_{B} T^{2} C$. Recognizing that the heat capacitance scales as
\begin{equation}
    C =  \gamma V_{\rm TES} T\,,
\end{equation}
where $\gamma$ is the specific heat coefficient for the appropriate metal, and $V_{\rm TES}$ is the volume of the TES, we find that the energy resolution scales as

\begin{equation}
    \sigma_{\rm E} \propto \sqrt{V_{\rm TES}\; T_{c,{\rm TES}}^3}\,.
\label{eq:TESScaling}
\end{equation}

In Table~\ref{tab:ExistingTES} we list the measured energy sensitivity of three of the most sensitive TES bolometers/calorimeters that exist today, along with their physical dimensions and operating temperatures $T_{c,{\rm TES}}$. What is immediately clear is that none of these devices have attempted to minimize both the TES volume and $T_{c,{\rm TES}}$ concurrently, and thus substantial sensitivity increases are conceptually possible. For an estimate of the potential gains, we have scaled these devices using Eq.~\eqref{eq:TESScaling} to the proposed TES geometry and operating temperature for our quasiparticle collection detector, which would be six 9~mK tungsten TES in parallel with dimensions of $1\;\mu{\rm m}\times 24\; \mu{\rm m} \times 35\;{\rm nm}$ each, for a total TES volume of  4.2~$\mu {\rm m}^{3}$ (see Table~\ref{tab:APQPDevice}). The resulting scaled energy resolutions $\sigma_{\rm E}^{\rm scale}$ are given in the right-most column of Table~\ref{tab:ExistingTES}. As is evident, ${\cal O}({\rm meV})$ energy sensitivities seem feasible.

\begin{table}[t]
\begin{center}
\begin{tabular}{ |c|c|c|c|c|c|c|c|}
\hline\hline \rule{0pt}{1.2em}
 {TES}              & $T_c$     & Volume                                &Bias Power         & Power Noise                           &$\tau_{\rm eff}$   & $\sigma_{\rm E}^{\rm measured}$   & $\sigma_{\rm E}^{\rm scale}$ \\
                    &  [mK]     &  [$\mu {\rm m}\times\mu {\rm m}\times{\rm nm}$]   &[W]                    & $\sqrt{S_{\rm{p,tot}}(0)}$ [W/$\sqrt{{\rm Hz}}$]  & [$\mu$s]  & [meV]                     &  [meV]  \cr \hline\hline
 W~\cite{NIST}          & 125   &$25\times 25\times 35$                     &$2.1\times 10^{-13}$   & $5 \times 10^{-18}$                   & 15            & 120                   & 1.1\\
 \hline
 Ti~\cite{Karasik}      & 50        &$6\times 0.4\times 56$                     &$5.8\times 10^{-17}$   & $2.97\times 10^{-20}$                 &           & 47                        & 22\\
                    & 100   &                                       &$2.6\times 10^{-15}$   & $4.2\times 10^{-19}$                  &           & 47                        & 7.8\\
\hline
 MoCu~\cite{SAFARI} & 110.6 &$100\times 100\times 200$                  &$8.9\times 10^{-15}$   & $4.2\times 10^{-19}$                  & 12700     & 295.4                     & 0.3\\
\hline\hline
\end{tabular}
\caption{Specifications and measured performance of three existing TES single photon calorimeters/bolometers. Energy sensitivity estimates for the TES design used in the quasiparticle collection device (Table~\ref{tab:APQPDevice}) are scaled from each device using the temperature and volume scalings of Eq.~\eqref{eq:TESScaling}. For the bolometer of Ref.~\cite{SAFARI}, energy resolution is estimated as the power noise multiplied by $\sqrt{\tau_{\rm eff}}$, where $\tau_{\rm eff}$ is the sensor fall-time.  For Ref.~\cite{Karasik}, energy sensitivity scalings are estimated for the device with $T_{c} =$100~mK as well as for the B-Field suppressed value of 50~mK.}
\label{tab:ExistingTES}
\end{center}
\end{table}

Unfortunately, improvements in TES sensitivity to low energy recoils via Eq.~\eqref{eq:TESScaling} are naturally accompanied by increased sensitivity to environmental noise. Since the thermal power flow between the TES and the bath scales with a power of $n_{\rm Tb} \sim5$, the bias power required to keep the TES within transition also scales as $T_{c}^{n_{\rm Tb}}$. To give a sense of scale, the TES for our proposed quasiparticle detector is estimated to have a bias power of $8.3\times 10^{-20}$~W, nearly 3 orders of magnitude smaller than that of current devices (Table~\ref{tab:ExistingTES}). Ideally, the bias power is predominantly supplied by the TES readout electronics, but this is certainly not necessarily the case.  Vibrations from cooling machinery (such as pulse tube cryocoolers and turbo pumps for dry dilution refrigerator systems, or 1K pot vibrations in wet systems) could dissipate power within the TES. Likewise, thermal radiation from poorly shielded higher temperature stages could be absorbed by the TES.  Finally, electromagnetic interference (even beyond the sensor bandwidth) can be coupled into the TES via the wiring. In summary, the constraints on DC environmental power loading are $\times10^{3}$ more strict than levels currently achieved.

Both electromagnetic interference and vibrational environmental noise sources will naturally have fluctuating components within the TES sensor bandwidth as well,  and thus as one decreases the fundamental thermal fluctuations between the TES and the bath, these sources could begin to dominate and suppress the TES sensitivity. Roughly, a decrease in environmental power noise by a factor of 50 from levels achieved today are required in order to meet the required TES performance specifications.

An automatic benefit of operating a TES at the low temperatures of the proposed detectors, is that the sensor fall-times (which are $\propto C/G$) naturally become very long (sensor bandwidths become very short) and so match the very long excitation-collection timescales that are envisioned, of ${\cal O}(10\;{\rm msec})$. Thus, the problem of bandwidth mismatch between the TES sensor and the excitation collection time, which degrades the energy sensitivity of current SuperCDMS~\cite{MCP_Thesis} and CRESST detectors~\cite{MCP_BandwidthMismatch}, is naturally suppressed.

Device specifications and estimated performance for both the proposed quasiparticle and athermal phonon excitation detectors are shown in Table~\ref{tab:APQPDevice}. In the table, we have assumed that the detector trigger threshold is 6 times the estimated detector baseline energy resolution, $\sigma_{\rm E\, D}$. Of course, $\sigma_{\rm E\,D}$ is just the TES baseline resolution ($\sigma_{\rm E\,TES}$) divided by the efficiency factors for collecting and concentrating excitation energy in the TES that were discussed in the text.

\begin{table} [th!]
\centering
\small
\begin{tabular}{|  l  |  l  |  l  | l |}
\hline\hline
                			&                                           				& {\bf Quasiparticle Detector}              			& {\bf Athermal Phonon Detector} \\
\hline\hline
                			& Number of Detectors                           		&  750                              						& 750\\
\hline\hline
                			&                                           				& Aluminum Absorber                 				& Tantalum Absorber \\
\hline\hline
                			& Absorber Volume                                        	& $5\times5\times5$ mm$^{3}$            			& $5\times5\times5$ mm$^{3}$\\
                        		& Excitation Scattering Length                    	& $>$ 5 mm  ($>$ 2 mm \cite{AlScattering})  		& $>$ 5 mm \\
                        		& Excitation Lifetime                               		& 20 ms ($>$ 2 ms \cite{0811.1961})         			& 1.2 ms \\
                        		&                                           				&                                   						& (1250 surface bounces) \\
 $f_{\rm cascade}$    & Fraction of Recoil Energy in                    	& $\sim$ 60\%                           					& $\sim$ 95\% \\
                        		& ~~~ Excitation System                         		&                                   						& (all QP have recombined \cite{0811.1961}) \\
                        		& Characteristic Group Velocity                      	& $\sim 2 \times10^{-3}$                    			& $10^{-5}$\\
\hline\hline
                			&                                           				& Tungsten QP Collector                     			& Aluminum Phonon Collector \\
\hline\hline
$A_{\rm collect}$	& Total Area of All Collection  				& $12\times 225\;\mu$m$^2$              			& $2 \times 0.21 \rm{mm}^2$\\
				&~~~ Fins on a Detector 					&										& \\
$h_{\rm collect}$   	& Thickness of Collection Fins             		&  $\sim$150 nm                     					& $\sim$ 900 nm \\
$f_{\rm trap}$      	& Excitation Trapping Fraction                 		&  0.1                              						& 0.5 \cite{MCP_Thesis} \\
$\tau_{\rm collect}$   & Excitation Collection Time                        	& 3 ms                                					& 700 $\mu$s \\
$f_{\rm collect}$   	& Excitation Collection Efficiency                  	&  87\%                             						& 63\% \\
$f_{\rm E\,Remain}$  & Fraction of Potential Energy                           & $\sim$ 0.90                           					& 0.60 $\times$ 0.65\\
                			& ~~~ Remaining After Collection                    	&                                   						& \\
\hline\hline
                			&                                          				& Tungsten TES                              				& Tungsten TES \\
\hline\hline
                			& Number of TES per detector                           & 6                                     					& 2 \\
$V_{\rm TES}$         	& Total Volume of all TES                           	&  $6\!\times 1\mu{\rm m} \! \times\! 20\mu{\rm m} \! \times \! 35{\rm nm}$ & $2 \! \times \! 1\mu{\rm m}\!\times\! 20\mu{\rm m} \! \times \! 35{\rm nm}$  \\
				&~~~ on a detector						&										&\\
$T_{c}$         		& Transition Temperature                            	&  9 mK                                 					&  9 mK\\
$C_{\rm TES}$       	& Heat Capacity                                 		&  $1.0\times10^{-17}$ J/K              				&  $4.0\times10^{-18}$ J/K \\
$\alpha$            	& Dimensionless Sensitivity                         	&  30                                   					& 30 \\
                			& Bias Power                                    			&   $7.0\times10^{-20}$ W              				&   $2.8\times10^{-20}$ W \\
$\sqrt{S_{\rm{p, tot}}(0)}$ & Total Power Noise                             	&   $4.4\times10^{-22}$ W/$\sqrt{\rm{Hz}}$  		&   $2.8\times10^{-22}$ W/$\sqrt{\rm{Hz}}$ \\
$\tau_{\rm eff}$        & Sensor Fall-Time                             			& 10 ms                             					& 10 ms    \\
                			& Collector to TES Efficiency                       	& 1                                 						& 0.74 \\
$\sigma_{\rm E\,TES}$    & TES Energy Resolution                     	&  0.3 meV                          					& 0.2 meV \\
$\sigma_{\rm E\,D}$  & Detector Recoil Resolution                        	&  0.6 meV                          					& 0.7 meV \\
				& ~=$\sigma_{\rm E\,TES}/(f_{\rm E\,Remain} f_{\rm collect} f_{\rm cascade})$&						&\\
 				& Energy Threshold (6 $\sigma_{\rm E\,D}$)       &  3.9 meV                          					& 4.2 meV \\
\hline\hline
\end{tabular}
\caption{ Specifications and estimated performance for both the quasiparticle and athermal phonon detectors. }
\label{tab:APQPDevice}
\end{table}

\subsection{Backgrounds}

\begin{figure}
\begin{center}
\includegraphics[width=0.75\textwidth]{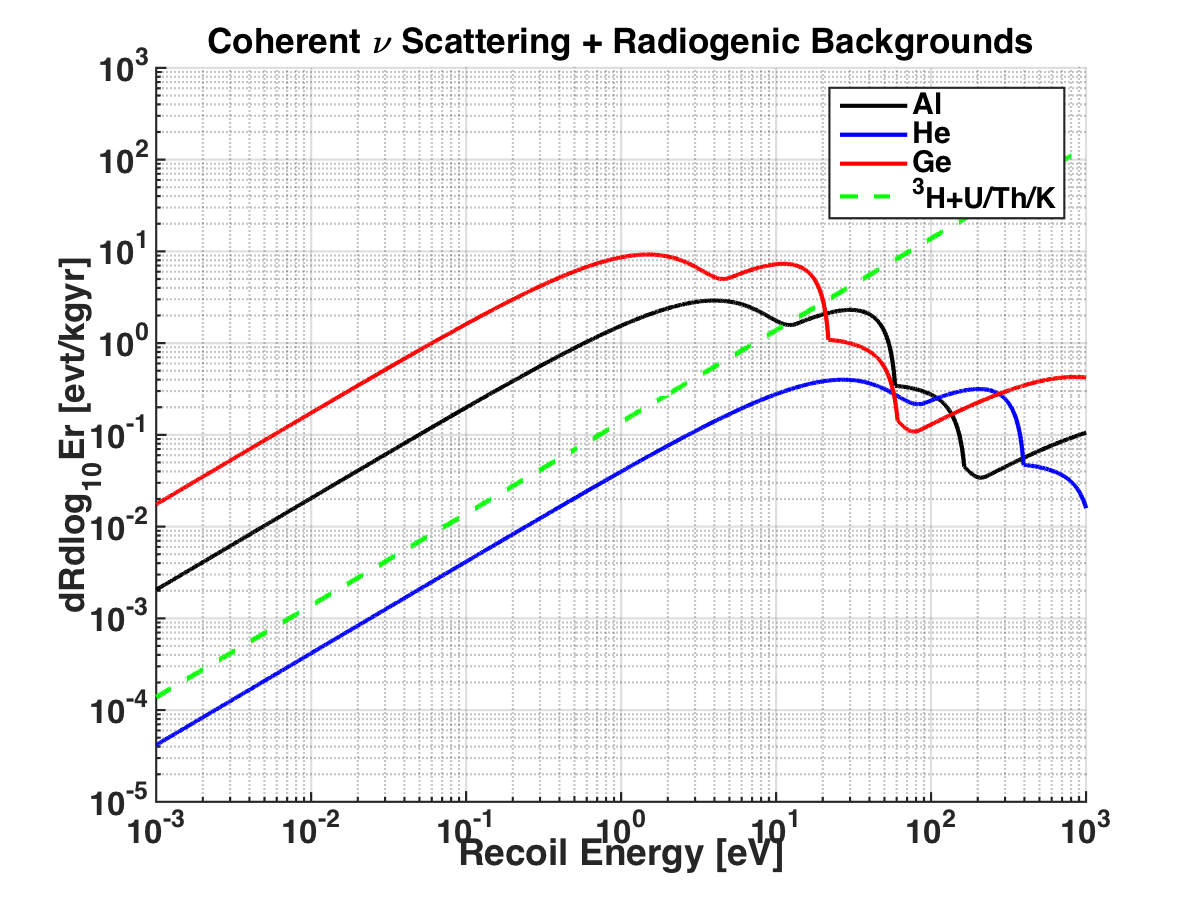}
\caption{Differential rate in units of $dR_\nu/d\log_{10}E_D$ for the solar neutrino coherent nuclear scattering background on various target nuclei as well as the expected radiogenic background from cosmogenic $^{3}$H spallation of the absorber during fabrication and from U/Th/K contamination of the SuperCDMS SNOLAB cryostat.}
\label{fig:NeutrinoBackground}
\end{center}
\end{figure}

Solar neutrinos are an irreducible background, with rate (per unit mass per unit time)
\begin{eqnarray}
 \label{eq:RateNeutrino}
E_D\frac{dR_{\nu}}{d E_D}=\int d E_\nu  E_D\frac{d\langle
n_T\sigma_\nu \rangle}{d E_D}\frac{1}{\rho} F_{\nu}\,,
\end{eqnarray}
where $F_\nu$ is the neutrino flux and $n_T$ the number density of the target.  As we focus on low energy
depositions in the detector, the dominant contribution comes from
$pp$ neutrinos~\cite{Bahcall:1987jc,Bahcall:1997eg} scattering on
nuclei.  This assumes that an ${\cal O}(1)$ fraction of the energy
deposited in nuclear recoils is converted into quasiparticles in the
detector.  The rate is shown in Fig.~\ref{fig:NeutrinoBackground},
for a few sample nuclei.  We find that for an aluminum target, the
integrated neutrino background for a kg$\cdot$year is less than
1~event for nuclear recoils between 1~meV and 1~eV, and is 3~events
for nuclear recoils between 10~meV and 10~eV.  We include the 3
background events where relevant in extracting DM limits,
accordingly.

The expected U/Th/K compton background of 13~event/keV/kg/year for Si detectors (a material with similar stopping power to Al) within the proposed SuperCDMS SNOLAB cryostat, plus the background due to the beta decay of $^{3}$H produced via cosmogenic spallation during detector fabrication at sea level (60 days assuming 125~atoms/kg/day production) are also shown in Fig.~\ref{fig:NeutrinoBackground} and found to be sub-dominant~\cite{SuperCDMS_CDR}. At first glance this might seem surprising because minimization of radiogenic backgrounds is the primary design driver in high mass dark matter direction detection. However, there are two reasons why radiogenic backgrounds are of secondary importance for light mass dark matter detection. First, the low energy coherent neutrino scattering background from $pp$ neutrinos is much larger than the background produced by atmospheric neutrinos within the high mass dark matter region of interest. Secondly, all of the radiogenic backgrounds (comptons, $^{210}$Pb decay products, $^{3}$H) have characteristic energy scales which are much larger than the light mass dark matter region of interest ($<$10 eV) and thus there is very little overlap between radiogenic backgrounds and light mass dark matter recoil signals.

\section{Dark matter scattering in a Fermi-degenerate medium}\label{sec:pauli}

Having established superconducting detector designs capable of
reaching meV energies, we now must establish DM scattering rates.
For the metal target studied here, we are interested in DM
scattering off the valence electrons, which, as previously described, are characterized by Fermi statistics, with typical
Fermi velocity $v_F \sim 10^{-2}$.
As the metal drops into a superconducting state at low temperature,
a $\sim$~meV gap opens up above the Fermi surface, blocking
DM-electron scattering for energy depositions below this gap.
For energy deposits well above the gap, the
scattering is simply characterized by allowable momentum
configurations of DM-electron scattering that are consistent with Fermi
statistics and Pauli blocking.   As the energy deposits drop and approach the gap,
an additional factor that takes into account the
presence of the superconducting gap, a so-called coherence
factor---similar to a form factor---kicks in~\cite{Tinkham}; this
factor depends on energy and has an effect only near threshold.
Since the energy thresholds we consider are always above the 0.3 meV
aluminum superconducting gap, we neglect the coherence factor in what follows.  We thus
approximate the electrons in the superconducting target as a free
Fermi-degenerate gas.

The most important property of the Fermi-degenerate metal or gas to properly
incorporate is the phase space suppression of Pauli blocking --- in a
Fermi-degenerate medium, the DM must deposit enough energy to knock
an electron out of the Fermi sea and into the continuum above the
Fermi surface.  We closely follow the discussion in
Ref.~\cite{Reddy:1997yr}, and reformulate their calculations for
non-relativistic DM-electron scattering.  We denote the 4-momentum
of DM initial and final states by $P_1$ and $P_3$, the initial and
final states of the electron by $P_2$ and $P_4$, and the momentum
transfer $q=(E_D,{\bf q})$. The scattering rate for a DM particle can be estimated via
\begin{eqnarray} \label{eq:response}
\langle n_e\sigma v_{\rm rel}\rangle&=&\int\frac{d^3
p_3}{(2\pi)^3}\frac{\langle |{\cal M}|^2\rangle}{16 E_1 E_2 E_3 E_4}\ S(E_D,{\bf q})\nonumber\\
S(E_D,{\bf q})&=&2\int\frac{d^3 p_2}{(2\pi)^3}\frac{d^3
p_4}{(2\pi)^3}(2\pi)^4\delta^4(P_1+P_2-P_3-P_4)f_2(E_2)(1-f_4(E_4))\,,
\end{eqnarray}
with $E_D$ the deposited energy, $\langle |{\cal M}|^2\rangle$ the squared scattering matrix element summed and averaged over spins, and $f_i(E_i)=\left[ 1+{\rm exp}\left(\frac{E_i-\mu_i}{T}\right)\right]^{-1}$ is the
Fermi-Dirac distribution of the electrons at temperature $T$.
$S(E_D,{\bf q})$ characterizes the Pauli blocking effects of the process at hand. If, for instance,
the scattering converts an electron to a different final state particle
which does not exhibit Pauli blocking, $(1-f_4(E_4))$ should be dropped in $S(E_D,{\bf q})$, and the integrals $d^3 p_3$ and $d^3 p_4$ in Eq.~(\ref{eq:response}) reduce to the ordinary 2-to-2 scattering phase space integral. Analytically,
$S(E_D,{\bf q})$ is found to be
\begin{eqnarray}
 \label{eq:ResponseExp}
S(E_D,{\bf q})=\frac{m_e^2 T}{\pi |{\bf q}|}\left[\frac{z}{1-e^{-z}}\left(1+\frac{\xi}{z}\right)\right]\,,
\end{eqnarray}
where
\begin{eqnarray}
 \label{eq:ResponseDetail}
z&=&\frac{E_D}{T}\,,\nonumber\\
\xi&=&\textrm{ln}\left[\frac{1+{\rm exp}[(e_- - \mu)/T]}{1+{\rm exp}[(e_- +E_D -
\mu)/T]}\right]\,,\nonumber\\
e_- &=& \frac{(E_D-|{\bf q}|^2/2m_e)^2}{|{\bf q}|^2/2m_e}\,,
\end{eqnarray}
and $\mu$ is the chemical potential, identified as $E_F$ at zero temperature.

In the limit of $T\rightarrow 0$, we have $z\rightarrow +\infty$ and
$\xi\rightarrow 0$, yielding
\begin{eqnarray}
 \label{eq:ResponseApprox}
S(E_D,{\bf q})\simeq \frac{m_e^2 E_D}{\pi |{\bf q}|}\;\Theta(|{\bf q}| v_F - |E_D|)\,,
\end{eqnarray}
with $\Theta$ the Heaviside theta function. We note that this limit
is only valid when both $e_-<\mu$ and $(e_-+E_D)<\mu$ (the small
region where only one inequality is satisfied is unimportant for the
rate estimation).
In what follows we compute the rate numerically using the full
Eqs.~\eqref{eq:ResponseExp} and~\eqref{eq:ResponseDetail} at
temperature much lower than the gap, in order to capture the entire
kinematic range properly.

Converting $d^3 p_3$ to energy $E_D$ and momentum transfer ${\bf q}$,
\begin{eqnarray}
 \label{eq:Jacobi}
d^3p_3=dE_D d|{\bf q}| \frac{2\pi |{\bf q}| (E_1-E_D)}{p_1 }\,,
\end{eqnarray}
the total interaction rate of DM-electron scattering, per unit mass per unit time, is
\begin{eqnarray}
 \label{eq:RateDM}
E_D\frac{dR_{DM}}{d E_D}=\int d v_X f_{\rm MB}(v_X)
E_D\frac{d\langle n_e\sigma v_{\rm rel}\rangle}{d
E_D}\frac{1}{\rho}\frac{\rho_{X}}{m_X}\,.
\end{eqnarray}
Here $f_{\rm MB}$ is the velocity distribution of DM, which we take
to be a modified Maxwell-Boltzmann distribution \cite{Smith:2006ym},
\begin{eqnarray}
 \label{eq:VelocityDist}
f_{\rm MB}(v_X)=\frac{4\pi
v_X^2}{N_E}e^{-v_X^2/v_{\rm esc}^2}\Theta(v_{\rm esc}-v_X)\,,
\end{eqnarray}
with the normalization factor
\begin{eqnarray}
 \label{eq:Normalization}
N_E=\left(erf(z)-\frac{2z e^{-z^2}}{\sqrt{\pi}}\right)\pi^{3/2}v_0^3\,,
\end{eqnarray}
with $z=v_{\rm esc}/v_0$. We use rms velocity $v_0=220 \textrm{km}/\textrm{s}$ and cut-off at the escape velocity
$v_{\rm esc}=500\; \textrm{km}/\textrm{s}$.
$\rho$ is the mass density of the detector material, $\rho_{X}/m_X$ the DM
local number density, with the DM mass density $\rho_{X}=0.3\
\textrm{GeV}/\textrm{cm}^3$.
A typical Fermi velocity is $v_F ={\cal O}(10^3)\; {\rm km/s}\gg v_{\rm esc}$, leading to $v_{\rm rel}\simeq v_F$ in Eq.~(\ref{eq:RateDM}).

As already discussed in Section~\ref{ssec:principle}, when the DM is lighter than the electron, there is always an electron
configuration in the Fermi sea that can fully stop the DM, and the energy cutoff is determined by the incoming kinetic energy of the
DM, yielding $E_D$ ranging from 0 to $\frac{1}{2}\mu_{X} v_{\rm
esc}^2$. In contrast, when the DM is heavier than the electron, no electron can fully stop
it, and the maximal deposited energy in this case is
$E_D^{\rm max}=\frac{1}{2}m_e[(v_F+2v_{\rm esc})^2-v_F^2]$.

\begin{figure}
\begin{center}
\includegraphics[width=0.65\textwidth]{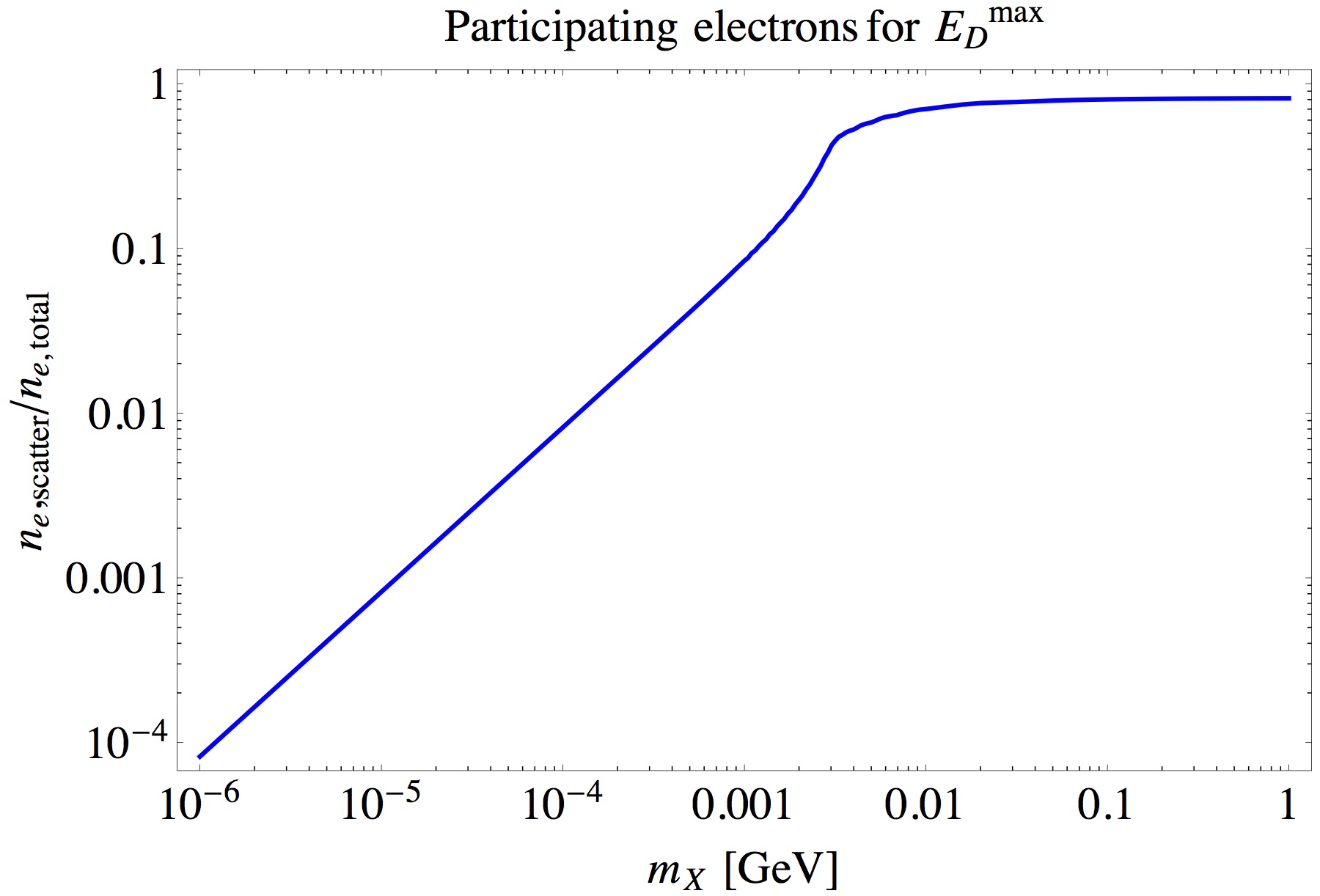}
\caption{The fraction of active electrons participating in the
scattering with maximal deposited energy, as a function of DM mass.
}\label{fig:frac}
\end{center}
\end{figure}

Intuitively, one expects the Pauli blocking effect to suppress the
total rate by $\sim E_D/E_F$, as this indicates the relative size of
the shell of electrons available for energy $E_D$ deposited in the scattering. We find that the complete
Pauli blocking computation is well captured by this naive estimation
when $E_D$ is small compared to Fermi energy.  As $E_D$ approaches $E_F$, more of the electrons can participate in the scattering, and the effect of  Pauli blocking is suppressed. As an illustration, we study the behavior of the fraction of
electrons participating in the scattering when the deposited energy is maximal, which is directly related to the DM mass. The fraction of participating electrons is plotted in Fig.~\ref{fig:frac}, and the numerical result agrees very well with the expectation: When DM is light, the maximal energy deposition is
small, and the fraction of active electrons grows linearly with DM mass.
When the DM mass is ${\cal O}({\rm MeV})$, the DM kinetic energy is comparable to Fermi
energy, and the linear growth approximation fails.

The scattering cross section between DM $X$ and free electrons with
a mediator $\phi$ is given by
\beq\label{eq:sigDD}
\sigma_{\rm scatter} = \frac{16\pi \alpha_e \alpha_X}{(m_\phi^2+{\bf q}^2)^2}\mu_{eX}^2\,,
\eeq
where $\alpha_i\equiv g_i^2/(4\pi)$, $g_i$ is the coupling of $\phi$
to $i=e,X$, $\mu_{eX}$ is the reduced mass of the DM-electron
system, and ${\bf q}$ is the three-momentum transfer in the process,
determined by the kinematics of the detection process. (Here we keep only
the contributions from the 3-momentum transfer $\bf q$ since the
energy transfer in a $t$-channel non-relativistic
scattering is much smaller.) This scattering cross section is related to the matrix
element squared in Eq.~\eqref{eq:response} through
\beq
\sigma_{\rm scatter}=\frac{\langle |{\cal M}|^2\rangle}{16\pi E_1 E_2 E_3 E_4}\mu_{eX}^2\,.
\eeq
We define two related reference cross sections, $\tilde \sigma_{\rm DD}$, corresponding to the light and heavy mediator regimes:
\beq\label{eq:sigDDdef}
\tilde \sigma_{\rm DD}^{\rm light} &=& \frac{16\pi \alpha_e \alpha_X}{q_{\rm ref}^4}\mu_{eX}^2\,,\quad q_{\rm ref} \equiv \mu_{eX} v_X\,,\nonumber\\
\tilde \sigma_{\rm DD}^{\rm heavy} &=& \frac{16\pi \alpha_e \alpha_X}{m_\phi^4}\mu_{eX}^2\,,
\eeq
where $v_X\sim 10^{-3}$ is the DM velocity. In the above, the
reference momentum $q_{\rm ref}$ is chosen for convenience as a
typical momentum exchange. It is worth noting that for very light DM that deposits $E_D$ energy, the momentum transfer in the process can be larger than this $q_{\rm ref}$ by a factor of $v_F/v_X$.
In-medium effects of massless mediators, which can alter Eq.~\eqref{eq:sigDD}, will be addressed when relevant.

\begin{figure*}[t!]
\begin{center}
\includegraphics[width=0.65\textwidth]{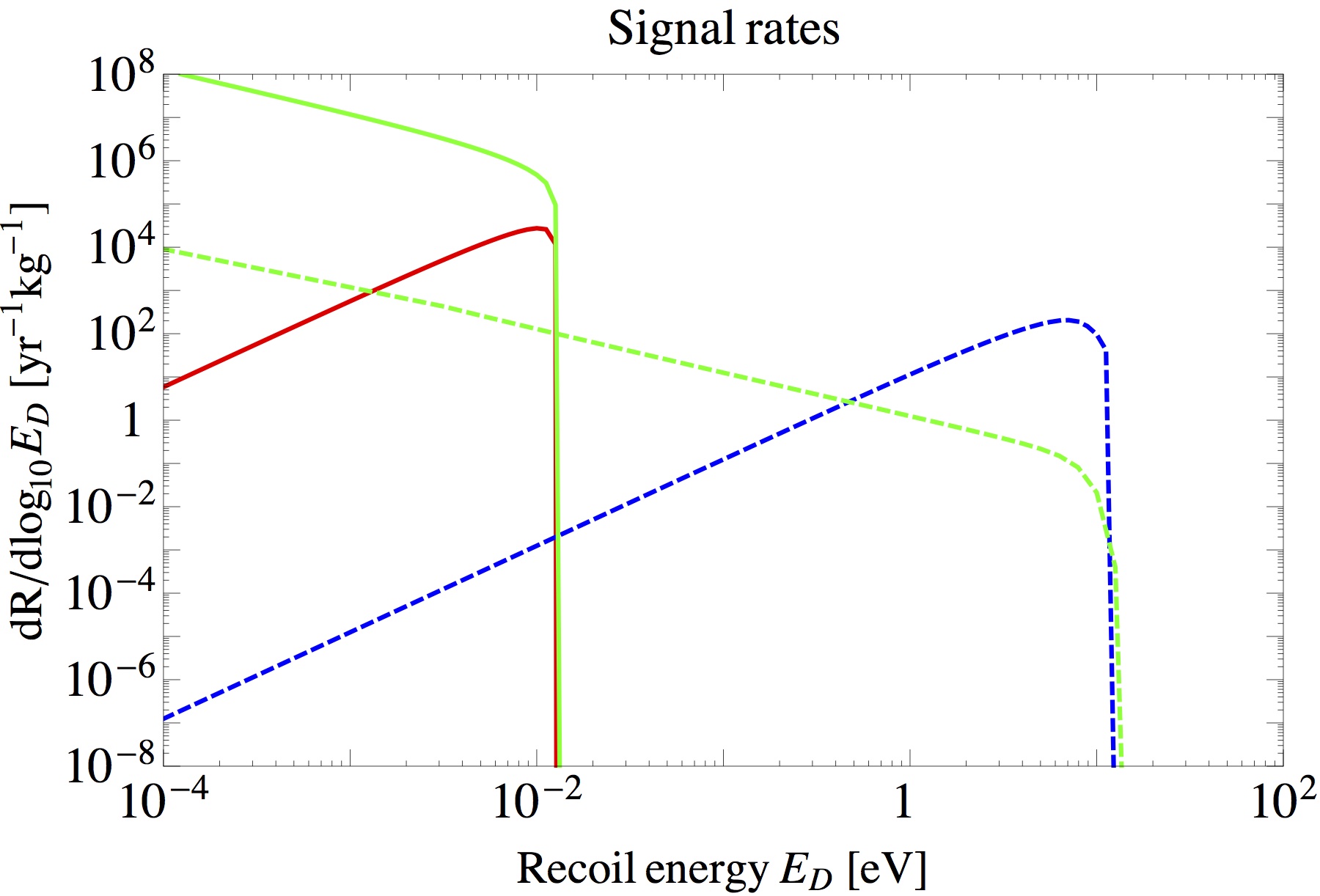}
\caption{ \label{fig:rate}
 Signal rates per kg$\cdot$year, for several benchmark
points of ($m_\phi, m_X, \alpha_X, g_e)$ = $(10\;\mu {\rm eV}, 10\; {\rm keV}, 5\times10^{-14}, 3\times10^{-9})$ [{\bf solid green}],  $(10\; \mu {\rm eV}, 100\; {\rm MeV}, 5\times 10^{-8}, 3\times 10^{-12})$ [{\bf dashed green}], $(1\;{\rm MeV}, 10\;{\rm keV}, 0.1, 3\times 10^{-6})$ [{\bf solid red}], and $(100\;{\rm MeV}, 100\;{\rm MeV}, 0.1, 3\times 10^{-5})$ [{\bf dashed blue}]. We use the Fermi energy of aluminum, $E_F=11.7$~eV.
The green [red and blue] curves correspond to a particular DM mass along the
same-colored curve in the top [bottom] panel of
Fig.~\ref{fig:DDlightS}.
 }
\end{center}
\end{figure*}

To establish a sense of a number of events expected, in Fig.~\ref{fig:rate} we plot the differential rate per kg$\cdot$year as a function of deposited energy $E_D$, for several benchmark points. The behavior of the curves can be readily understood. As we have seen, the maximum energy deposition is controlled by the DM mass, independently from the mediator. When the mediator is heavy compared to the momentum transfer of the process, the rate is peaked at high energy depositions. This is because, when the energy deposition is well below Fermi energy, the larger the energy deposits, the larger the fraction of participating electrons. For very light mediators, the rate is dominated by the minimal momentum transfer in the process, which is controlled by the detector energy threshold.

\section{Cosmological, astrophysical and terrestrial constraints}\label{sec:constraint}

Having established the DM interaction rate in our proposed detectors, we now consider the types of constraints such
DM is subject to in order to determine whether DM candidates consistent with all constraints are within reach.

\subsection{Self-interactions}\label{ssec:SIDM}

DM self-interactions bound $g_X$ via a constraint on the scattering
cross section weighted by the momentum transfer,
\beq
\sigma_T=\int d\Omega_* \frac{d\sigma}{d\Omega_*}(1-{\rm cos}\theta_*)\,.
\eeq
In the Born regime, where $\alpha_X m_X \ll m_\phi$, the analytic
perturbative result~\cite{Tulin:2012wi,Tulin:2013teo} for attractive and
repulsive forces is
\beq\label{eq:sigTborn} \sigma_T^{\rm Born} =
\frac{8\pi\alpha_X^2}{m_X^2
v^4}\left[\log(1+R^2)-\frac{R^2}{1+R^2}\right]\,,\quad R\equiv m_X
v/m_\phi\,, \eeq
which reduces in the heavy mediator limit of $m_\phi\gg m_X v$, as expected, to the contact operator form,
\beq\label{eq:sigTheavy}
\sigma_T^{\rm {\rm heavy}}\approx
\frac{4\pi \alpha_X^2 m_X^2}{m_\phi^4}\,.
\eeq
For very light mediators in the classical regime, where $m_X v \gg m_\phi$, the solution to the classical equations of motion in repulsive and attractive potentials  (see {\it e.g.}~\cite{Tulin:2012wi} and references therein) reduces to
\beq\label{eq:sigTlight}
\sigma_T^{{\rm light}} \approx
\frac{16\pi \;\alpha_X^2}{v^4 m_X^2}\ln \beta^{-1}\,, \quad \beta = \frac{2m_\phi\alpha_X}{m_X
v^2} \ll 1\,,
\eeq
in the limit of $\beta \ll 1$, which will always be applicable to our light (but massive) mediator case.  Here we have taken Dirac DM with interactions via a vector or scalar mediator; a Majorana or real scalar DM particle would have a factor of 4 larger scattering cross-section.

Bullet-cluster constraints~\cite{Clowe:2003tk,Markevitch:2003at,Randall:2007ph} along with recent simulations which reanalyze the constraints from halo shapes~\cite{Rocha:2012jg,Peter:2012jh}, limit the DM self-interaction cross section
to be roughly
\beq\label{eq:sigTlim}
\frac{\sigma_T}{m_X} \lsim 1-10 \ \units{cm^2/g}\,,
\eeq
depending on the relevant velocity; further details can be found
{\it e.g.} in Ref.~\cite{Kaplinghat:2015aga}. The self-interaction
constraints will be most relevant when discussing light mediators,
where the transfer cross section is proportional to $1/v^4$. In
order to be conservative, in later discussions in
Section~\ref{sec:models}, we use $\sigma_T\lsim 10$~cm$^2/$g with
$v\sim 10^{-4}$ to impose an upper bound on $g_X$. For the very
light mediator regime, this roughly translates to requiring
\beq
\left(\alpha_X\right)_{\rm SIDM}^{\rm light} \lsim 4\times10^{-17} \left(\frac{m_X}{{\rm keV}}\right)^{3/2}\left(\frac{v}{10^{-4}}\right)^2\left(\frac{58}{\ln\beta^{-1}}\right)^{1/2}\,,\quad \beta = \frac{2m_\phi\alpha_X}{m_X v^2}\,,
\eeq
where $\ln\beta^{-1}$ varies by a factor of at most a few in the region of interest.

\subsection{Kinetic decoupling}\label{ssec:kindec}

The couplings of a light mediator $m_\phi\lsim$~eV to DM and to
electrons are constrained by CMB measurements. This is because if DM
is in kinetic equilibrium with the the photon-baryon plasma during
the recombination epoch, DM density fluctuations can be washed out
via Silk damping~\cite{Silk:1967kq} and the baryon acoustic peak
structure can be altered.

Closely following Ref.~\cite{McDermott:2010pa}, we require the relaxation rate of energy transfer is slower than the expansion rate of the universe: %
\beq \label{eq:kindec}
\Gamma_p = \left.\sum_{b = e,p} \frac{8 \sqrt{2
\pi} n_b \alpha_X \alpha_b \mu_{bX}^{1/2}}{3 m_X T^{3/2}} \ln
\left[\frac{3 T \lambda_{\rm cut}}{\sqrt{\alpha_b \alpha_X}} \right] \right|_{T=\hat T} \lsim\left. H\right|_{T=\hat T}\,, \eeq
where $\mu_{bX}$ is the reduced mass of the baryon-DM,
and $\lambda_{\rm cut}$ is the screening length for the baryon plasma; for massive mediators, this is set by $1/m_\phi$, while for photon exchange, it corresponds to the Debye screening mass, $\lambda_D=\sqrt{T/(4\alpha_{\rm EM} \pi n_e)}$ with $n_e$ the electron density. For light mediators, Eq.~\eqref{eq:kindec} is to be evaluated at the time of recombination, $\hat T=T_{\rm rec}\simeq 0.26$~eV. For heavy mediators having $m_\phi^2\gsim T_{\rm rec}m_X$, ensuring the interaction is out of equilibrium at recombination is not enough -- we must ensure that DM be decoupled from the plasma when the momentum transfer is of order the mediator mass, leading to $\hat T = m_\phi^2/(4\mu_{bX})$ in Eq.~\eqref{eq:kindec} above. In the discussion in Section~\ref{sec:models} we compare this constraint on $\alpha_e\alpha_X$ against others and require the strongest of them holds. Of course, kinetic decoupling need only be enforced when the DM and/or the mediator have masses below a few MeV, and at temperatures below a few MeV. The constraint is most relevant for light mediators, where it roughly reads
\beq
\left(\alpha_X \alpha_e\right)_{\rm kin.\;dec.}^{\rm light}\lsim 10^{-19} \left(\frac{m_X /\sum_{b=e,p}\sqrt{\mu_{bX}}}{{\rm keV}^{1/2}}\right)\left(\frac{50}{\ln}\right)\,,
\eeq
within ${\cal O}(1)$ factors, for the entire mass ranges of interest to us, where we use at recombination $n_{e,{\rm rec}} = n_{p,{\rm rec}} \simeq 2.4\times 10^{-39}\; {\rm GeV}^{-3}$.

We note that in the above we have considered kinetic decoupling
between the DM and the SM particles it scatters with via exchange
of the mediator.  In the case of a light mediator, the mediator itself (rather than the DM) may be brought into thermal equilibrium via Compton-like processes; requiring the mediator is out of equilibrium during BBN results in a constraint on $g_e$ weaker than those from stars we consider next.

\subsection{Stellar emission}\label{ssec:stellar}

If DM scatters off electrons via the exchange of a mediator, the
mediator may be emitted from stellar objects and generate excess
cooling. A variety of emission processes can occur~--- amongst them
Compton-like processes, bremstrahlung off electrons, Primakoff
emission and plasmon decay/conversion. We now summarize the
relatively model-independent constraints on the mediator coupling to electrons $g_e$ that
are generally applicable to a scalar mediator; the relevant constraints on a
non-kinetically mixed vector differ by an ${\cal O}(1)$ factor~\cite{Raffelt:1996wa}.

For mediator masses beneath a keV, we find that the dominant stellar constraint comes from Horizontal Branch (HB) star cooling.
The bremstrahlung process imposes the strongest bound~\cite{Raffelt:1996wa,Grifols:1988fv},
\beq\label{eq:brem}
g_e^{\rm brem}\lsim1.3\times 10^{-14}\quad [{\rm HB}]\,,
\eeq
which is stronger by roughly a factor of 3 than that from the Compton-like process~\cite{Raffelt:1996wa}.
We have verified that the loop-level induced coupling of $\phi$ to photons, which generates additional emission via the Primakoff process, yields a constraint on $g_e$ that is $\sim 3$ orders of magnitude weaker than Eq.~\eqref{eq:brem}.

The above constraint Eq.~\eqref{eq:brem} is,
in principle, applicable to all mediators with mass beneath the typical stellar temperature of $\sim 10$~keV.   We emphasize, however, that this is not always the case; in particular models, such as the kinetically mixed light hidden photon, the constraint on the coupling $g_e$ is much weaker.  This will be addressed separately in
Section~\ref{ssec:hiddenU1}.

For mediator masses exceeding several ten's of keV, the relevant
stellar environment becomes the hotter supernovae, though in these
dense objects, stellar constraints can be lifted due to trapping
effects. There can be model-dependence in which process controls
trapping, but decays of the mediator allow electron couplings of
order $g_e\gsim 10^{-6}$ at the very least (see {\it e.g.}
Ref.~\cite{Dent:2012mx}) for mediator masses above a few hundred
keV.  Here terrestrial experiments can play a vital role.

\subsection{Terrestrial constraints}\label{ssec:terres}

Terrestrial experiments are complementary probes to the electron-coupling constraints considered thus far.
Restricting to processes that do not require coupling to additional SM particles beyond electrons, the relevant experimental bounds come from measurements of the anomalous magnetic
moment of the electron $(g-2)_e$~\cite{Hanneke:2008tm,Bouchendira:2010es,Aoyama:2012wj},
beam dump experiments such as E137~\cite{Batell:2014mga},
and low energy $e^+e^-$ machines such as BaBar~\cite{Aubert:2008as,Essig:2013vha}.

When the mediator is heavier than twice the electron mass, it can decay visibly to a pair of electrons; if it is also heavier than twice the DM mass, invisible decays open up as well. Although a broad region of parameter space for $m_\phi\gsim {\rm MeV}$ decaying either visibly or invisibly is constrained by terrestrial searches, a viable window of parameter space with fairly large couplings opens up even for mediator masses as light as a few MeV. This is due to supernova trapping effects, in a similar way to the case of the well-studied kinetically mixed hidden photon.  For a sense of the size of the allowed couplings, a $m_\phi\sim10$~MeV [$100$~MeV] mediator with couplings $\alpha_e\sim{\rm few}\times 10^{-6}$ [few$\times10^{-5}$] and $\alpha_X\sim 0.1$ can evade all terrestrial and astrophysical constraints (see {\it e.g.} Ref.~\cite{Batell:2014mga}). For sub-MeV DM and mediator couplings this large, vanilla DM models may be brought into thermal equilibrium with the photon plasma and affect $N_{\rm eff}$ and/or BBN~\cite{Boehm:2013jpa}.  If the DM is a real scalar, even if the DM is brought into thermal equilibrium, it remains consistent with BBN constraints.  These constraints can also be evaded in `non-vanilla' models, for instance by varying the couplings or the masses of the system with temperature.
A full exploration of such models will be detailed in a separate publication~\cite{future}.

\section{Models and results}\label{sec:models}

Having discussed the generic bounds that are relevant for
constraining the allowed size of the direct-detection cross section,
we now move to describing several concrete models and their results.
Scalar and vector mediators are addressed in
Section~\ref{ssec:scalar}; a kinetically mixed hidden photon (including in-medium effects) is treated in
Section~\ref{ssec:hiddenU1}; milli-charged DM is considered in
Section~\ref{ssec:mili}; and dipole-interacting DM is detailed in
Section~\ref{ssec:dipole}.

\subsection{Scalar and vector mediator}\label{ssec:scalar}

We begin by considering a real scalar mediator $\phi$, described by the potential
\beq
{\cal L}_{\rm scalar} = -\frac{1}{2}m_\phi^2 \phi^2+ g_e \phi \bar e e + g_X \phi \bar X X\,.
\eeq
The DM-electron t-channel scattering for a vector mediator, such as
a $U(1)_{B-L}$ vector boson, with tree level couplings to electrons
and DM, are the same as those obtained for the scalar case, in the non-relativistic limit. We consider mediators both lighter and heavier than the
momentum transfer involved in the process.

\begin{figure*}[h!]
\begin{center}
\includegraphics[width=0.63\textwidth]{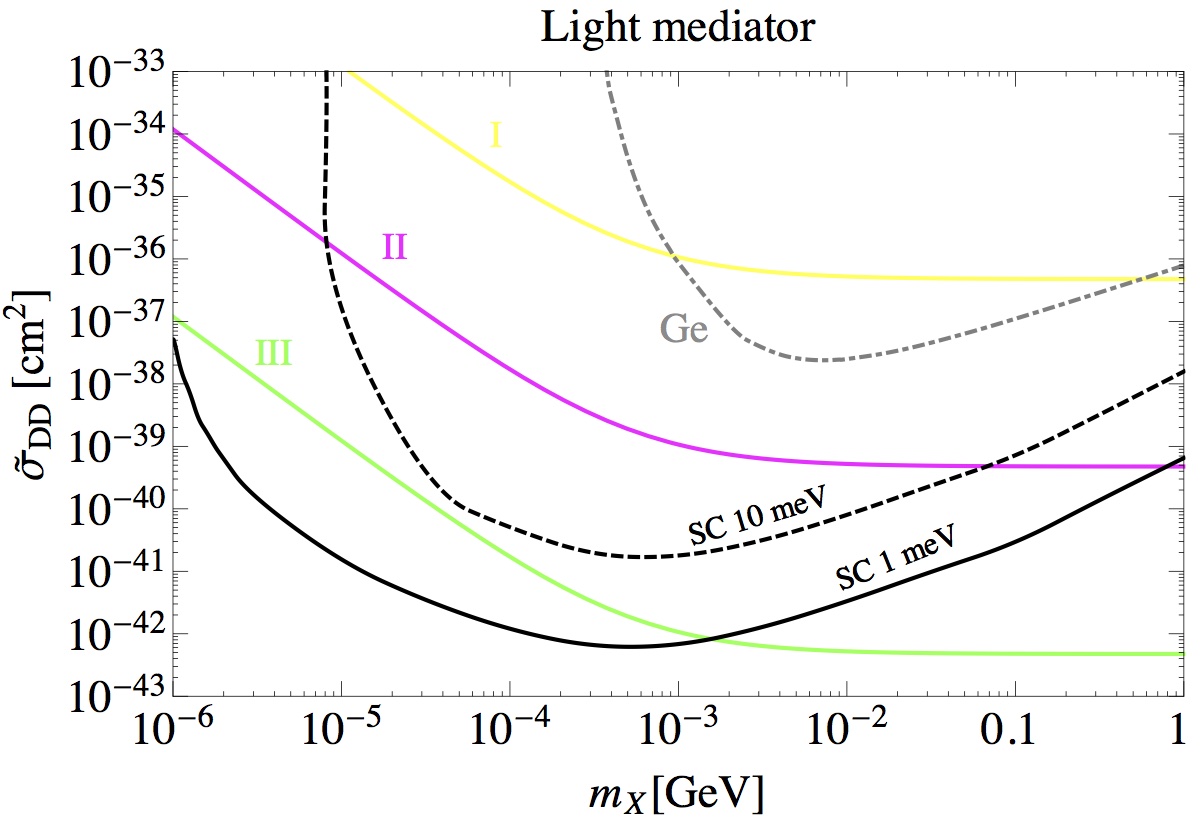}\\
~\\
\includegraphics[width=0.63\textwidth]{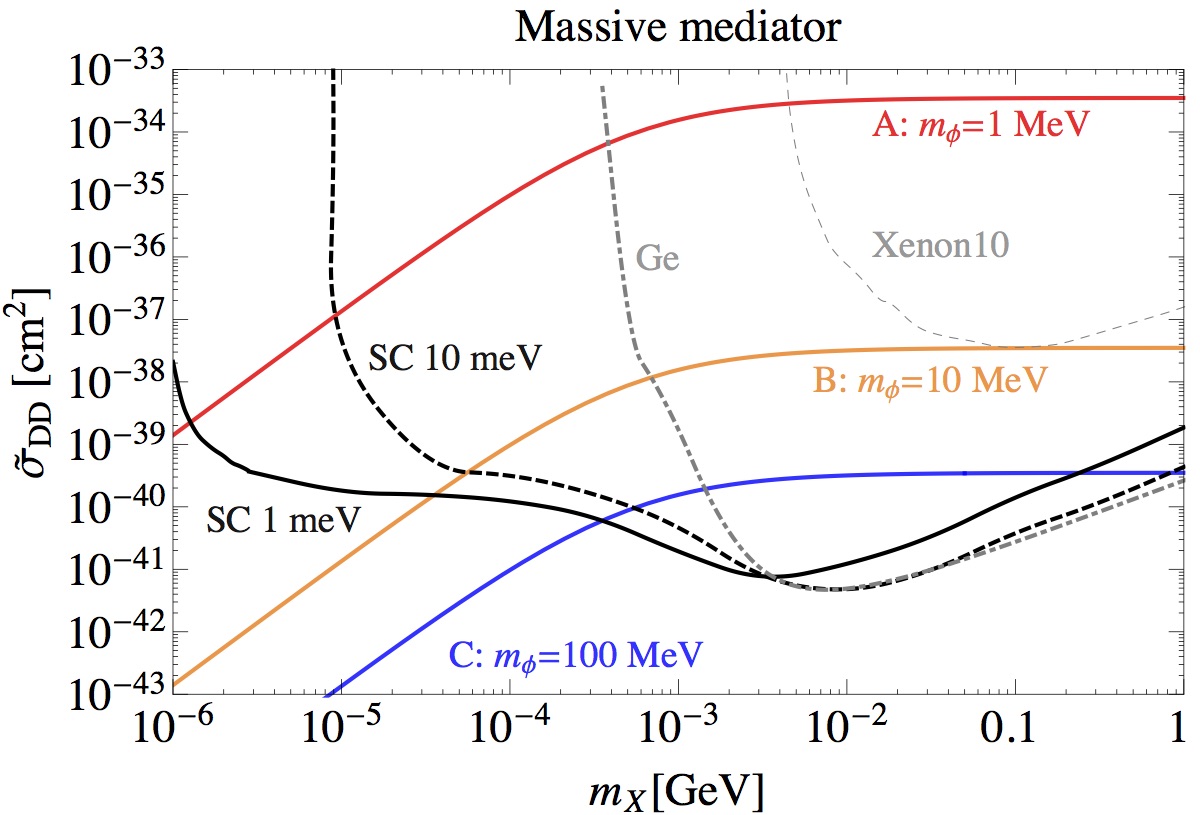}
\caption{ \label{fig:DDlightS} {\bf Top:} Direct detection cross
section, Eq.~\eqref{eq:sigDDdef}, for light DM scattering off
electrons via a scalar or (non kinetically mixed) vector mediator, for several benchmarks.
These are {\bf I:} $\alpha_X=10^{-15}, \alpha_e=10^{-12}$;
{\bf II:} $\alpha_X=\alpha_e=10^{-15}$; and {\bf III:}
$\alpha_X=10^{-15}, \alpha_e=10^{-18}$. These depicted parameters
obey bounds from self-interactions and decoupling at recombination
for $m_\phi\lsim {\rm eV}$, though stellar emission (and BBN
considerations for vectors) may place strong constraints; see text
for details. {\bf Bottom:} Direct detection cross section between
light DM and electrons, for several benchmarks of heavy mediators.
These are {\bf A:} $m_\phi=1$~MeV, $g_e=10^{-5}e$, $\alpha_X=0.1$;
{\bf B:} $m_\phi=10$~MeV, $g_e=10^{-5}e$, $\alpha_X=0.1$; and {\bf
C:} $m_\phi=100$~MeV, $g_e=10^{-4}e$, $\alpha_X=0.1$. These depicted
parameters obey all terrestrial and stellar-cooling
constraints, though sub-MeV DM interacting with SM through a
massive mediator may be strongly constrained by BBN; see text for
details. The Xenon10 electron-ionization data
bounds~\cite{Essig:2012yx} are plotted in thin dashed gray. {\bf In
both panels}, the black solid (dashed) curve depicts the sensitivity
reach of the proposed superconducting aluminum devices, for a
detector sensitivity to recoil energies between 1~meV$-$1~eV
(10~meV$-$10~eV), with a kg$\cdot$year of exposure. We have included only the solar neutrino background in our estimate. For comparison,
the gray dot-dashed curve depicts the expected sensitivity utilizing
electron ionization in a germanium target as obtained in
Ref.~\cite{Essig:2015cda}. }
\end{center}
\end{figure*}

{\bf Light scalar/vector mediator.}
The self-interaction constraints of Eq.~\eqref{eq:sigTlim} can be combined together with the kinetic decoupling
requirements of Eq.~\eqref{eq:kindec} and the stellar bounds of Eq.~\eqref{eq:brem} to
learn how large the scattering cross section of DM and electrons Eq.~\eqref{eq:sigDD} can be.
In the top panel of Fig.~\ref{fig:DDlightS} we plot $\tilde
\sigma_{\rm DD}^{\rm light}$ of Eq.~\eqref{eq:sigDDdef} for several
benchmark points, labeled {\bf I-III}, shown in solid colored
curves. As is evident, large cross sections can be obtained even for
very small couplings. This is due to the enhancement of the cross
section at low momentum transfer when the mediator is light, as shown in Fig.~\ref{fig:DDlightS}. The
presented benchmarks all obey self-interaction constraints and also
ensure that DM is kinetically decoupled through the time of
recombination for mediator masses $m_\phi\lsim$~eV. However, the depicted benchmarks may bring the mediator into equilibrium with the SM plasma via Compton-like processes; this is not a problem for a real scalar mediator, but a light vector mediator can then contribute too many degrees of freedom to $N_{\rm eff}$~\cite{Boehm:2013jpa}. Likewise, the stellar constraints of Eq.~\eqref{eq:brem} are 
very stringent. In vanilla models, we find that stellar cooling is too severe to allow
for a detectable rate for a light real scalar or vector mediator,
which constrains $g_e$ itself to be below the $10^{-14}$ level. We
learn that unless stellar bounds are somehow lifted, the direct
detection experiments considered in this paper will be unable to
probe scalar or vector mediators with masses below $\sim {\cal
O}(10)$~keV. Stellar and BBN constraints may be lifted, however, for
instance via having the coupling $g_e$ vary with environment, or via
trapping effects, in which case sizable direct detection cross
sections can be accommodated~\cite{future}. A kinetically mixed vector mediator can also lift stellar constraints, as will be discussed in Section~\ref{sssec:U1stellar}.

{\bf Heavy scalar/vector mediator.} Moving to a massive scalar or
vector, we focus on $m_\phi\gsim$~few~MeV. In the bottom panel
of Fig.~\ref{fig:DDlightS} we plot $\tilde \sigma_{\rm DD}^{\rm
heavy}$ of Eq.~\eqref{eq:sigDDdef} for several benchmark points,
labeled ${\bf A-C}$, which survive all terrestrial and stellar
cooling constraints, as outlined in Sections~\ref{ssec:stellar}
and~\ref{ssec:terres}. Fairly large couplings to electrons are
possible despite supernova constraints due to stellar trapping
effects, and beam dump constraints can be evaded by decaying
invisibly to additional dark-sector particles. As mentioned earlier,
for values of $\alpha_e$ and $\alpha_X$ as large as these benchmark
points, DM and/or the mediator can be brought into thermal
equilibrium with the SM plasma. BBN and Planck limits on the
effective number of relativistic degrees of freedom in equilibrium
$N_{\rm eff}$~\cite{Boehm:2013jpa} thus place constraints on, at
least, the simplest of such models. If the mediator is
heavy enough that it does not contribute to $N_{\rm eff}$ at BBN, if DM is
a real scalar, then $N_{\rm eff}$ constraints are trivially satisfied.  Even if a real scalar mediator, along with a real scalar DM, is in thermal equilibrium at BBN, $N_{\rm eff}$ constraints (which allow one additional fermion, or two additional real scalars) are still satisfied at 95\% C.L. 
For DM with more degrees of freedom, these bounds can potentially be
lifted; this is the case, for instance, if the couplings and/or the
masses of the particles involved evolve during the thermal history
of the universe. (This is much in the spirit of
Refs.~\cite{Anderson:1997un,Fardon:2003eh,Weiner:2005ac}.) Model
building efforts along these lines, both for relaxing BBN/$N_{\rm
eff}$ constraints for light or heavy mediators as well as lifting
stellar constraints for light mediators, are being pursued in detail
elsewhere~\cite{future}.

{\bf Reach.}
The $95\%$ expected sensitivity reach for a kg$\cdot$year of our proposed
superconducting aluminum experiment for light and heavy mediators is
depicted in the thick black curves of both panels of Fig.~\ref{fig:DDlightS}, with
the dashed [solid] curves showing the expected sensitivity with a 10~meV [1~meV] operating threshold.  Given that the heat sensors on the
detector are likely to have a somewhat limited dynamic range, we
also place an upper bound of 10 eV [1 eV] on the detectable energy. The depicted curves then correspond to 8.8 [3.7] events per kg$\cdot$year, taking into account the expected 3 [$<1$] neutrino background events as found in Section~\ref{sec:pauli}.
For completeness, we show the Xenon10 electron-ionization
bounds~\cite{Essig:2012yx} in the thin gray dashed curves (these are
absent in the top panel, as they are orders of magnitude weaker
than the displayed parameter space). We also show the projected
reach curves utilizing electron ionization techniques in a
semi-conductor germanium target (silicon performs similarly) as obtained in
Ref.~\cite{Essig:2015cda}, translated to $\tilde \sigma_{\rm DD}$ of
Eq.~\eqref{eq:sigDDdef}, shown in the thick gray dot-dashed curves.
For massive mediators and DM heavier than a few hundred keV, the
projected reach from a germanium/silicon target is comparable to than our
proposed detection method, while for lower DM masses, where electron
ionization techniques lose sensitivity, superconducting devices win.
Light mediators further demonstrate the strength of our proposed
detectors. When the mediator is light, superconductors can
out-perform electron-ionization techniques by several orders of
magnitude for dark masses above several hundred keV. Going to even
lighter masses, superconducting detectors are uniquely staged to
probe such super light DM.

\subsection{Kinetically mixed $U(1)_D$}\label{ssec:hiddenU1}

Next we study DM scattering with electrons through a kinetically
mixed dark $U(1)_D$.  We consider a hidden photon mediator $A'$ which is
kinetically mixed with the ordinary electromagnetic photon,
\beq
{\cal L}\supset -\frac{1}{4} F_{\mu\nu} F^{\mu\nu}
-\frac{1}{4}F'_{\mu\nu}F'^{\mu\nu}-\frac{\epsilon}{2}
F_{\mu\nu}F'^{\mu\nu}+\frac{m_{A'}^2}{2}A'^\mu A'_\mu + e J^\mu_{\rm
EM}A_\mu + g_X J^\mu_{\rm DM}A'_\mu\,,
\label{KineticLagrangian}
\eeq
with $F_{\mu\nu}$ ($F'_{\mu\nu}$) the (hidden) photon field strength
and $\epsilon$ the kinetic mixing parameter. $J_{\rm EM}^\mu$
and $J_{\rm DM}^\mu$ are the dark and electromagnetic currents, respectively.

Diagonalizing the kinetic terms and moving to the mass basis, the
hidden photon couples to the electromagnetic current of the SM with
strength $g_e=e\epsilon$. The mass of the
hidden photon $m_{A'}$ is obtained via a dark Higgs mechanism or Stuckelberg mechanism.
We note that we do not include the terms
involving the dark Higgs in the above Lagrangian. These will be relevant only when we
consider stellar cooling processes; they do not affect the
DM-electron scattering of the direct detection process. Indeed, the
direct-detection scattering rate is the same regardless of the mass
mechanism for the hidden $U(1)_D$, be it via a dark Higgs or the
Stuckelberg case.

\subsubsection{Photon propagator in medium}\label{sssec:prop}

In any model that can be written in the form of Eq.~\eqref{KineticLagrangian}, the size of the effective kinetic mixing parameter, $\epsilon_{\rm eff}$, is medium-dependent:
\beq
\epsilon_{\rm eff} = \epsilon\frac{q^2}{q^2-\Pi_{T,L}}\,,
\eeq
where here $q = (\omega,{\bf q})$ is the four-momentum transfer of a process and $\Pi_{T,L}$ is the in-medium polarization tensor, defined according to
\beq
\Pi^{\mu\nu} = \Pi_T \sum_{i=1,2} \epsilon_i^{T\mu} \epsilon_i^{T*\nu} + \Pi_L \epsilon^{L\mu}\epsilon^{L\nu}\,,
\label{PolarizationTensor}
\eeq
with $\epsilon^{T,L}$ the transverse and longitudinal polarization vectors:
\begin{eqnarray}\label{eq:eps_tensor}
\epsilon^L & = & \frac{1}{\sqrt{q^2}}(|{\bf q}|,\omega \frac{{\bf q}}{|\bf q|})\,, \\
\epsilon^T_{1,2} & = &\frac{1}{\sqrt{2}} (0,1,\pm i,0)\,.
\end{eqnarray}

Now the question is how to extract $\Pi_{L,T}$.   In the Appendix, we show using Maxwell's equations that, for a
non-magnetic medium,
\begin{eqnarray}
q^2(1-\tilde{n}^2) & = &  \Pi_L\,, \no\\
\omega^2(1-\tilde{n}^2) & = &  \Pi_T\,,
\label{PiLPiT}
\end{eqnarray}
where $\tilde{n} = n - i k$ is a complex index of refraction that is
related to the conductivity $\sigma$ and electric
permittivity $\varepsilon_r$ via~\cite{younglecturenotes}
\beq 1 - \varepsilon_r = 1 - \tilde{n}^2 = -\frac{i
\sigma}{ \omega} .
\label{SigmaL}
\eeq

\begin{figure*}[t!]
\begin{center}
\includegraphics[width=1\textwidth]{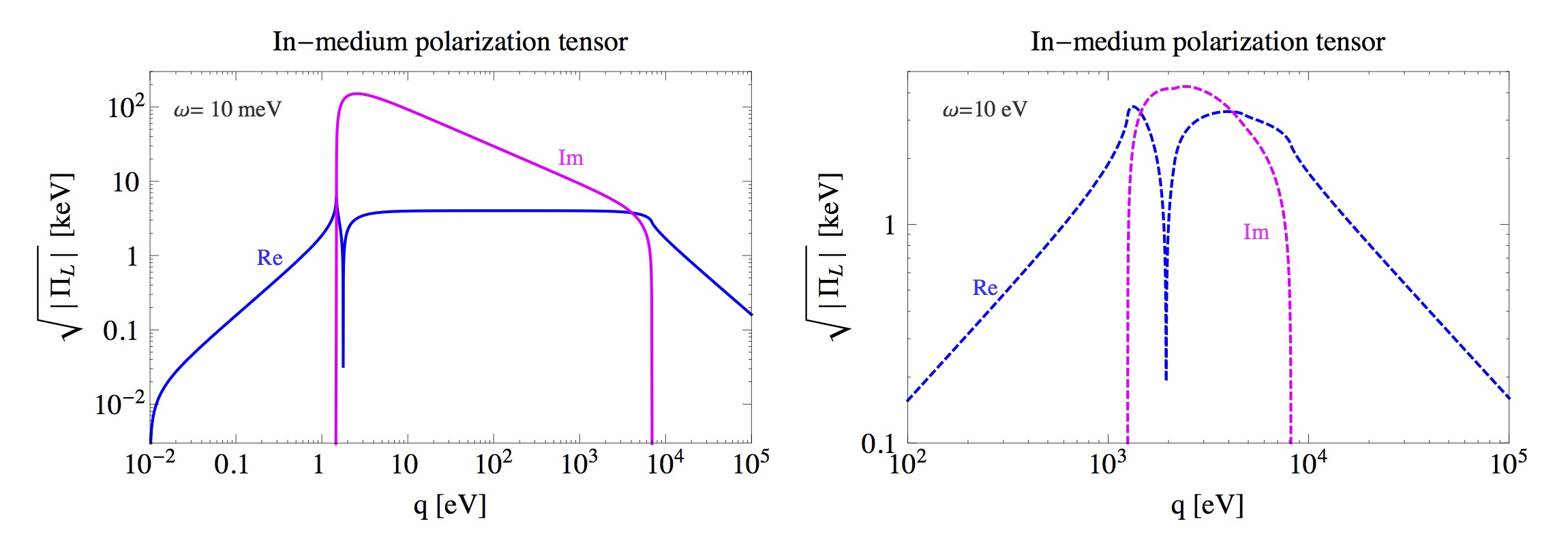}
\caption{ \label{fig:Pi} Real and imaginary parts of the in-medium
polarization tensor $\sqrt{\Pi_L}$ as a function of momentum
transfer, for deposited energies $\omega=10$~meV ({\bf left}) and
$\omega=10$~eV ({\bf right}). Here we use the Fermi energy of
aluminum, $E_F=11.7$~eV.
 }
\end{center}
\end{figure*}

The conductivity of a target metal, such as superconducting
aluminum, differs tremendously from that of an insulating target,
such as Helium. For Helium, the index of refraction is very
close to unity, implying that the photon mass in Helium is negligibly
small. In contrast, in a metal like aluminum, the electric
permittivity can be quite large and the photon mass appreciable. The
relative permittivity as a general function of $\omega$ and ${\bf q}$ is
given by~\cite{dressel}:
\begin{eqnarray}\label{ElectricPermittivity}
\varepsilon_r & = & 1 +
\frac{\lambda_{\rm TF}^2}{|{\bf q}|^2}\left\{\frac{1}{2} + \frac{p_F}{4 |{\bf q}|}\left[1 -
\left(\frac{|{\bf q}|}{2 p_F} - \frac{\omega}{|{\bf q}| v_F}\right)^2\right]\ln
\left\{\frac{\frac{|{\bf q}|}{2 p_F}-\frac{\omega}{|{\bf q}| v_F} + 1}{\frac{|{\bf q}|}{2
p_F}-\frac{\omega}{|{\bf q}| v_F} - 1}\right\} \right. \\ \nonumber
&& \left.\quad + \frac{p_F}{4 |{\bf q}|}\left[1-\left(\frac{|{\bf q}|}{2 p_F} + \frac{\omega}{|{\bf q}|
v_F}\right)^2\right]\ln \left\{\frac{\frac{|{\bf q}|}{2 p_F}+\frac{\omega}{|{\bf q}|
v_F} + 1}{\frac{|{\bf q}|}{2 p_F}+\frac{\omega}{|{\bf q}| v_F} - 1}\right\}\right\}\,,
\end{eqnarray}
where $\lambda_{\rm TF}^2 = 3 e^2 n_e/(2E_F)$ is the Thomas-Fermi screening
length.  For aluminum, $\lambda_{\rm TF} \simeq 4$~keV, taking $E_F = 11.7$ eV.  In Fig.~\ref{fig:Pi} we show the real and imaginary parts of $\sqrt\Pi$ as a
function of $|{\bf q}|$ with fixed values of $\omega$.  Note that the imaginary part is only non-zero in a limited range of $|{\bf q}|$ and $\omega$ where the kinematics allows a photon to be absorbed;  this corresponds to $\omega$ between $\omega = \frac{1}{2 m_e}(2 |{\bf q}| p_F + {\bf q}^2)$ and $\omega = \frac{1}{2 m_e}(2 |{\bf q}| p_F - {\bf q}^2)$ (or with an overall minus sign, depending on the choice of $\omega$).   As is evident, for typical ${\bf q} \sim v_F \omega$ the
effective photon mass in medium is approximately $\sim$~keV, on the order of the Thomas-Fermi screening length, implying
that for typical momentum transfers of order 10~eV for $m_X = 1
\mbox{ keV}$, the direct detection rate in metals is severely
limited.   Note that it is the Thomas-Fermi screening length (of order a few keV), and not the plasma mass (typically ${\cal O}(10\;{\rm eV}$)), which is the relevant screening parameter for scattering processes, where $q \gg \omega$; the plasma mass becomes the relevant screening mass for processes where $\omega \gg q$. We learn that for a kinetically mixed hidden photon
mediator, an insulating target is preferred.

We can now incorporate these in-medium effects and compute the
scattering cross-section for DM off of a nucleus or an electron via
the exchange of a dark $U(1)_D$.  The matrix element is given by
\begin{eqnarray}\label{Matrix}
\mathcal {M}=g_\chi \;e\;\bar u(p_3) \gamma^\mu u(p_1)\ G_{\mu\nu}(q)\
\bar u(p_4) \gamma^\nu u(p_2)\,,
\end{eqnarray}
where
\begin{eqnarray}
 \label{eq:propagator}
G_{\mu\nu}(q)=\frac{g_{\mu\alpha}-q_\mu
q_\alpha/q^2}{q^2-m_{A'}^2}\times\epsilon(q^2 g_{\alpha
\beta}-q_\alpha q_\beta) \times G_{{\rm IM},\beta\nu}\,.
\end{eqnarray}
Here $G_{{\rm IM},\beta\nu}$ is the in-medium photon propagator,
which can be parameterized in Lorentz gauge
as~\cite{Schmitt:2014eka}
\begin{eqnarray}
 \label{eq:propagatorD}
G_{{\rm
IM},\mu\nu}(q)=\frac{P_{L,\mu\nu}}{\Pi_L-q^2}+\frac{P_{T,\mu\nu}}{\Pi_T-q^2},
\end{eqnarray}
where the projection operators are
\begin{eqnarray}
 \label{eq:projection}
P_T^{00}&=&P_T^{0i}=P_T^{i0}=0\,,\nonumber\\
P_T^{ij}&=& \delta^{ij}-\hat q^i \hat q^j\,,\nonumber\\
P_L^{\mu\nu}&=&\frac{q^\mu q^\nu}{q^2}-g^{\mu\nu}-P_T^{\mu\nu}\,.
\end{eqnarray}

Utilizing the Ward identity, one finds that the second term of the
first factor in Eq.~\eqref{eq:propagator} vanishes.  Further, since
we are only interested in non-relativistic scattering between the DM
and electron, the zeroth components of the external momenta are much
larger than the spatial components. In the non-relativistic limit,
we find that the leading contribution comes from the longitudinal
component, with the transverse components suffering velocity
suppression. Thus in the following calculation, we keep only the
longitudinal component of the photon propagator, and use
\begin{eqnarray}
 \label{eq:propagatorLong}
G_{\rm IM}^{\mu\nu}&=&\frac{g^{\mu\nu}}{q^2(1-\Pi_L/q^2)}\nonumber\\
&=&\frac{g^{\mu\nu}}{q^2\left(1-\Pi_{00}/|{\bf q}|^2\right)}\,.
\end{eqnarray}
where we use the relation $\Pi_L=\frac{q^2}{|{\bf
q}|^2}\Pi_{00}$. Plugging this back to Eq.~\eqref{eq:propagator}, and
simplifying using the Ward identity, we find
\begin{eqnarray}
 \label{eq:propagator3}
G_{\mu\nu}(q)=\frac{\epsilon\
g_{\mu\nu}}{\left(q^2-m_{A'}^2\right)\left(1-\Pi_{00}/|{\bf
q}|^2\right)}\,.
\end{eqnarray}
Combining Eqs.~(\ref{Matrix}) and~(\ref{eq:propagator3}) we obtain
\begin{eqnarray}
 \label{eq:Matrix2}
\langle{|\mathcal {M}|^2}\rangle &\simeq& \frac{16 m_e^2 m_\chi^2
g_\chi^2
e^2\epsilon^2}{\left(q^2-m_{A'}^2\right)^2\left(1-\Pi_{00}/|{\bf
q}|^2\right)^2}\,,
\end{eqnarray}
where in the above we used the non-relativistic approximation, with $q=(\omega,{\bf q})=(p_1-p_3)$.

\begin{figure*}[t]
\begin{center}
\includegraphics[width=0.65\textwidth]{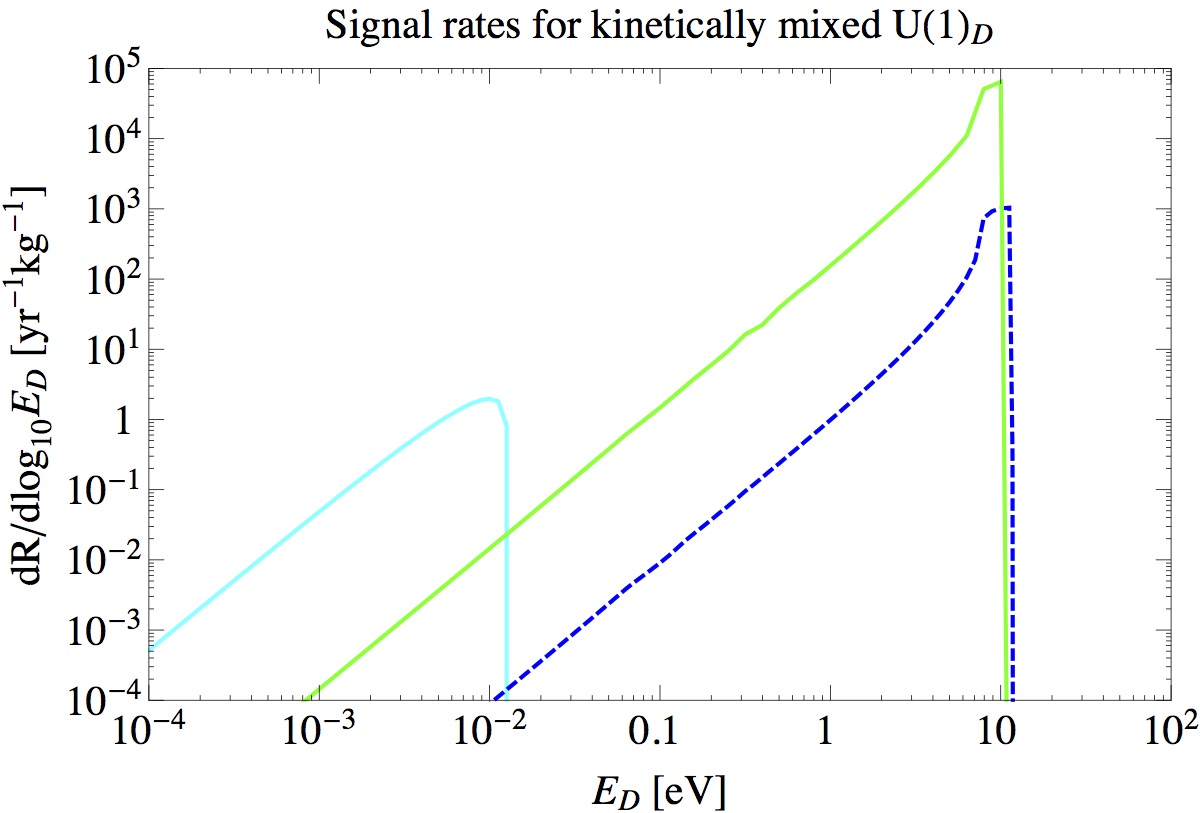}
\caption{ \label{fig:KINrate}
 Signal rates per kg$\cdot$year for a kinetically mixed hidden photon, for several benchmark
points of ($m_\phi, m_X, \alpha_X, g_e)=(10^{-14}\;{\rm eV}, 10\; {\rm
keV}, 2\times10^{-15}, 0.01)$~[{\bf solid cyan}],  $(0.1\;{\rm meV}, 100\; {\rm MeV}, 2\times 10^{-9}, 10^{-8})$~[{\bf solid green}], and $(100\;{\rm MeV}, 100\;{\rm MeV}, 0.1, 3\times 10^{-5})$~[{\bf dashed blue}]. We use the Fermi energy of aluminum,
$E_F=11.7$~eV. The solid cyan and green [dashed blue] curves
correspond to a particular DM mass along the same-colored curve in
the left [right] panel of Fig.~\ref{fig:DDcombKIN}.
 }
\end{center}
\end{figure*}

Utilizing Eqs.~\eqref{eq:response},~\eqref{eq:RateDM} and~\eqref{eq:Matrix2},
we can now compute the rate for an aluminum target. The differential rate per kg$\cdot$year as a function of deposited energy is given in Fig.~\ref{fig:KINrate}, for several benchmark points. (Note that for heavy mediator and very light DM, the rate is always substantially smaller than the depicted range; as a result we do not show a corresponding benchmark point.)
Comparing to Fig.~\ref{fig:rate}, we find as expected that the in-medium effects essentially modify the qualitative behavior of the light mediator into that of a massive one. The resulting projected reach of such a metal target will be reduced accordingly, increasing the desirability to find an insulating target with small gap.

\subsubsection{Kinetically mixed stellar constraints}\label{sssec:U1stellar}

An upper limit on the size of the direct detection cross section
arises due to constraints on the relevant couplings: DM
self-interactions constrain $g_X$; stellar cooling bounds the size
of $g_e$; and requiring kinetic decoupling of the DM and mediator
from the SM plasma such that CMB measurements are obeyed constrains
the combination of the two couplings. The self-interactions and
kinetic decoupling constraints presented in
Section~\ref{sec:constraint} directly apply here. Stellar emission
constraints for a light kinetically mixed hidden photon differ,
however, from those presented above, and are largely lifted, as
we now discuss.

For the hidden photon masses in our range of interest
$m_{A'}\lsim$~eV, stellar constraints have been worked out in detail in
the literature~\cite{An:2013yfc, An:2013yua}.
The dark photons are emitted from the sun and horizontal branch stars
either through plasmon resonance conversion,
or in association with photon decays, via a Higgstrahlung
process. The former process proceeds regardless of the origin of the hidden vector's mass, while the latter exists only in the case
of a dark Higgs mechanism. The crucial difference between these two
cases arises in the small-$m_{A'}$ limit: The Higgstrahlung process
does not vanish with vanishing mass, while
the plasmon resonance conversion scales with $\propto m_{A'}^2$ and
vanishes for a massless mediator~\cite{An:2013yua};
for further details, see Refs.~\cite{An:2013yfc, An:2013yua}.
In the small mass region, where $m_{A'}\ll \omega_p$ with
$\omega_p\sim 100$~keV the plasma frequency in the sun and horizontal
branch stars, the (direct) emission power of dark photons per volume is governed by the
emission of longitudinal modes of $A'$, and is proportional to
$\propto m_{A'}^2\omega_p^3\alpha_e$. The rate for the Higgstrahlung
process is governed in the small mass region by decays of transverse
photons, as they are more abundant than the longitudinal plasmons.
The total energy power density of dark radiation is then proportional to $\propto\omega_p^5 \alpha_e \alpha_X q_{H_D}^2$, with $q_{H_D}$ denoting the dark Higgs charge under the $U(1)_D$ (relative to the DM charge).
The resulting stellar constraints are found to be~\cite{An:2013yua}:
\beq\label{eq:stellarU1}
{\rm Higgstrahlung}&:&\quad \epsilon\;\left(\frac{q_{H_D} g_X}{0.1}\right)\lsim 8\times10^{-14} \quad [{\rm HB}]\,,\no\\
{\rm Resonance\ conversion}&:& \quad \epsilon \;\left(\frac{m_{A'}}{\rm
eV}\right)\lsim 4\times 10^{-12}\quad [{\rm Sun}]\,, \eeq
for mediator masses $10^{-5}\; {\rm eV}\lsim m_\phi\lsim {\rm eV}$
that we consider. For even lighter mediator masses, $\epsilon$ is
bound by photon-dark photon mixing through level-crossing in the
CMB~\cite{Jaeckel:2010ni}, as well as from the CROWS
experiment~\cite{Betz:2013dza,Graham:2014sha} and measurements of
deviations from Coulomb's law~\cite{Jaeckel:2010ni}; these are
lifted for $m_\phi\lsim 10^{-14}$~eV, where measurements of the
shape of the static magnetic field of Jupiter allows kinetic mixing
as large as ${\cal O}(10^{-2}-1)$ (see {\it e.g.}
Refs.~\cite{Jaeckel:2010ni,Graham:2014sha} for a summary of
constraints).

Combined with the self-interaction constraints on $\alpha_X$ and stellar emission constraints on the DM as well, one
can identify as a function of $m_X$ and $m_{A'}$ the strongest
constraints and place a bound on the combination $\alpha_e
\alpha_X$ which enters the direct detection cross section.

In the case of a dark Higgs mechanism, assuming similar dark-charges
of the dark Higgs and the DM, we find that for $m_{A'}$ below $\sim
10^{-5}$~eV, the cooling is dominated by the Higgstrahlung process
in the entire DM mass range of interest, despite the strong
suppression of $g_X$ from self-interactions. For 0.1~meV$\lsim
m_{A'}\lsim $~eV, plasmon resonance conversion dominates the cooling
for light DM masses until the Higgstrahlung process takes over; for
$m_{A'}=$~meV the turnover point is $m_X\sim 200$~keV, and increases
with increased $m_{A'}$. For $m_{A'}\gsim $~eV, combined with the
self-interaction bounds on $g_X$, plasmon resonance conversion is
most important for the entire $m_X$ range of interest. If a
hierarchy between the dark Higgs and DM charges is present, the
Higgstrahlung constraints can be relaxed accordingly. Likewise, in
the Stuckelberg case, only plasmon resonance conversion is relevant,
and as the hidden photon mass decreases, the stellar bounds on $A'$ are
weakened, at which point the other bounds mentioned above play a role.

We note that the analysis of Refs.~\cite{An:2013yfc, An:2013yua}
does not include effects of trapping and absorption in the relevant
stellar objects, which can, in principle, open up parameter space
above the constraints presented there. Taking into account
the low density in these stellar objects compared to that of
supernovae, however, we expect that trapping becomes important only for very
large kinetic mixing values where other (terrestrial)
observations already exclude the parameter space.

\subsubsection{Kinetically mixed results}\label{sssec:U1comb}

The direct detection cross section between electrons and DM through the
exchange of kinetically mixed hidden photon can now be constrained.
We take into account self-interactions via Eq.~\eqref{eq:sigTlim},
kinetic decoupling via Eq.~\eqref{eq:kindec} and the stellar bounds
via Eq.~\eqref{eq:stellarU1}. Due to the plasma effects of the photon
propagator, we choose to plot here the direct detection of
Eq.~\eqref{eq:sigDDdef}, times $(q_{\rm ref}/{\rm keV})^4$, namely
we plot
\beq\label{eq:sigDDhat}
\hat \sigma_{\rm DD}^{\rm light/heavy} \equiv \tilde \sigma_{\rm DD} ^{\rm light/heavy} \times \left(\frac{q_{\rm ref}}{\rm keV}\right)^4\,,
\eeq
where in the above we have taken the photon plasma mass $\Pi_L$ at a typical value of $\sim$~keV.  We consider separately the light and heavy mediator regimes.

\begin{figure}[t!]
\begin{center}
\includegraphics[width=0.65\textwidth]{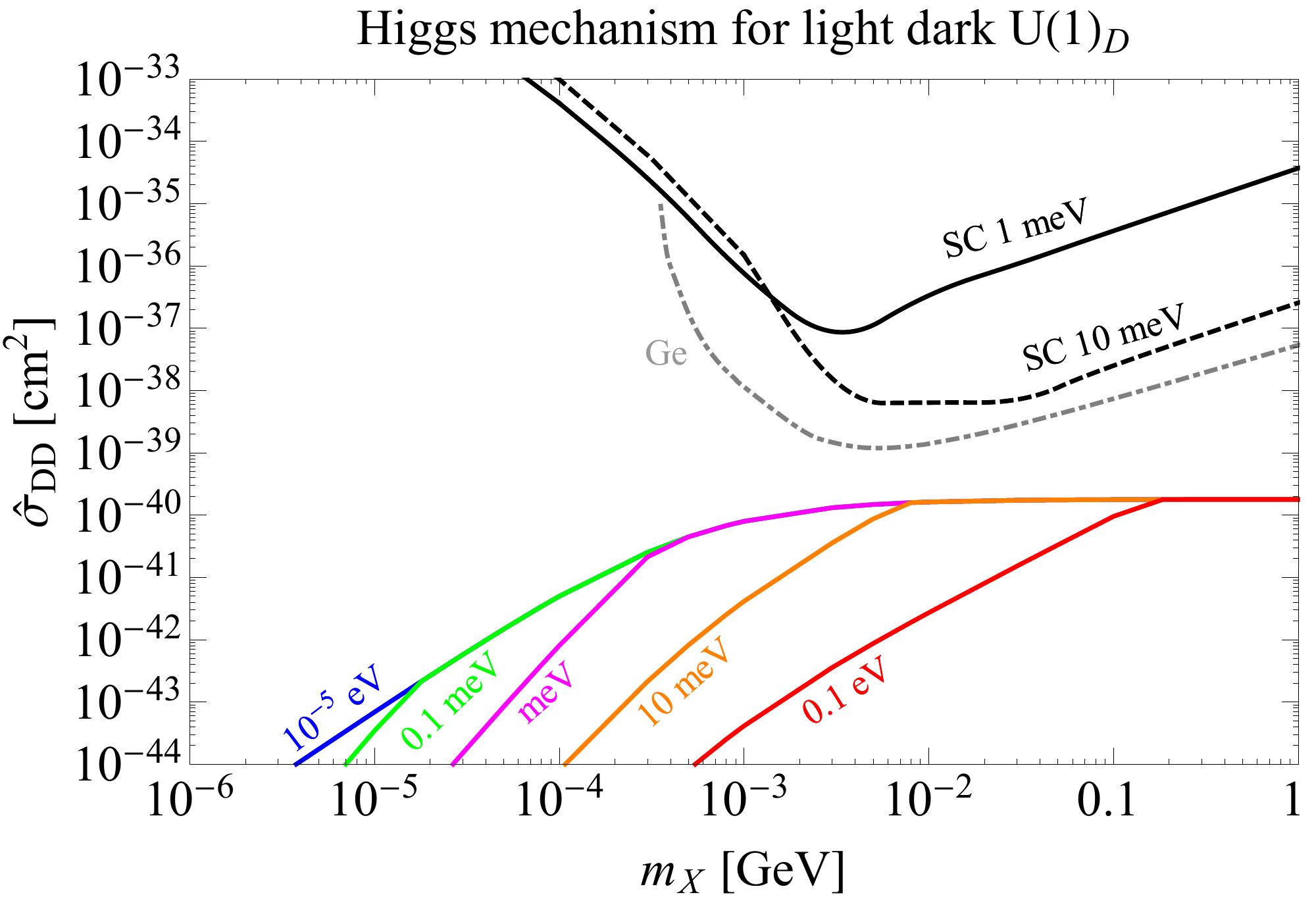}
\caption{Upper bounds on the direct detection cross section of
Eq.~\eqref{eq:sigDDhat} for light DM scattering off electrons, for a
light kinetically mixed hidden photon mediator obtaining its mass
via a dark Higgs mechanism, for a variety of different mediator
masses (solid colored curves). The expected reach of a
superconducting aluminum target with sensitivity to energies between
10~meV-10~eV and 1~meV-1~eV, as well as a germanium
target~\cite{Essig:2015cda}, is shown in thick dashed black, solid
black and dot-dashed gray, respectively.  We have included only the solar
neutrino background in our estimate.  The
in-medium effects of a metal target do not enable DM detection via a
superconducting metal in this case due to strong stellar constraints
on the relevant coupling; detection of these models would require an
insulating target.}\label{fig:DDhiggs}
\end{center}
\end{figure}

{\bf Light mediator.} The largest allowed direct-detection reference
cross section $\hat{\sigma}_{\rm DD}$ for the Higgs case with $q_{H_D}\sim 1$ is depicted in the solid colored curves of
Fig.~\ref{fig:DDhiggs}, for a variety of light mediator masses
$m_{A'}\lsim$~eV.
The kink in the curves as the mass of the DM increases is due to the
change in the stellar constraints as the dominant cooling mechanism
evolves (factoring in self-interaction constraints on $\alpha_X$)
from plasmon resonance conversion emitting the $A'$ to the Higgstrahlung process, as
detailed above.

\begin{figure*}[t!]
\begin{center}
\includegraphics[width=0.63\textwidth]{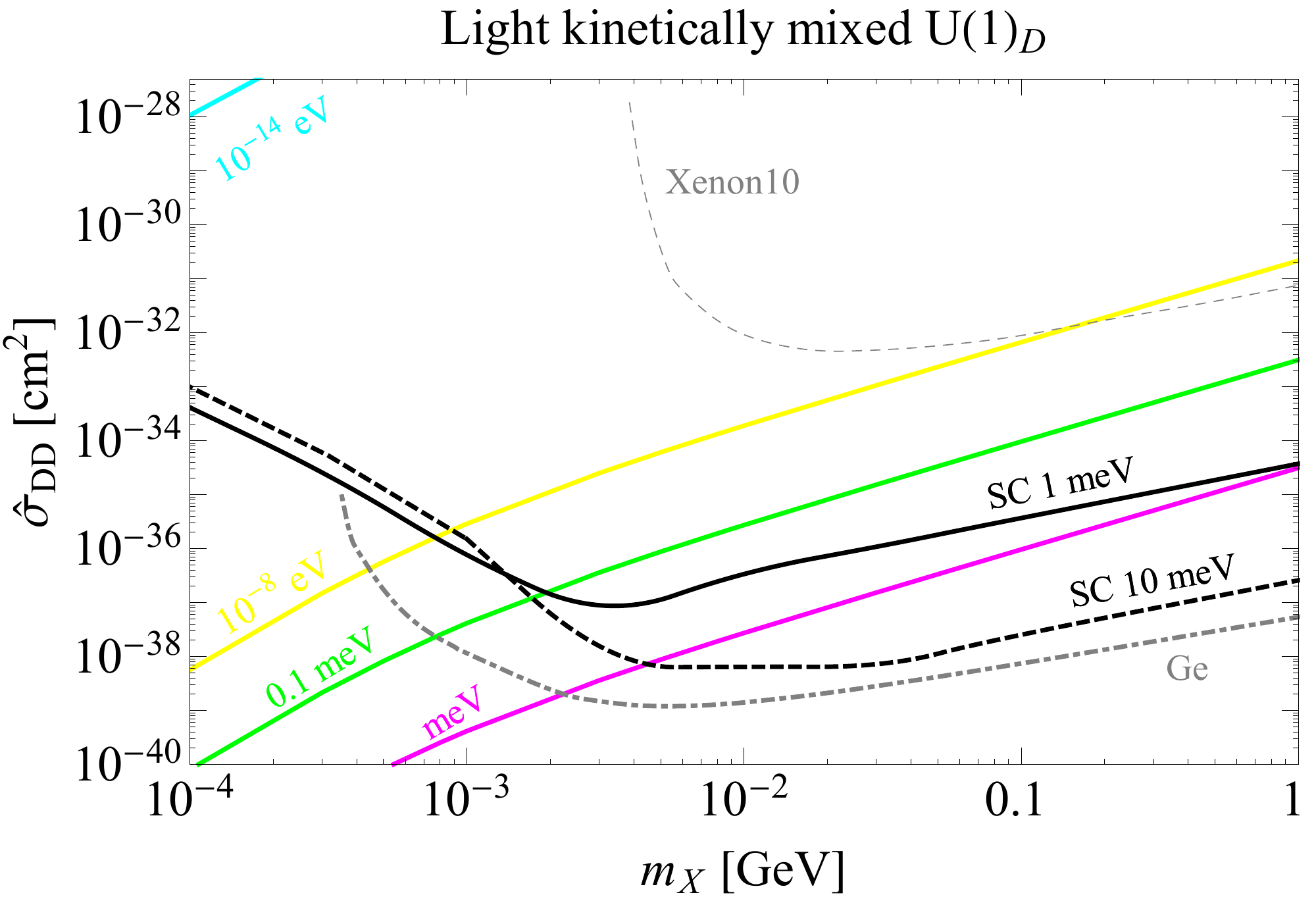}\\
~\\
\includegraphics[width=0.63\textwidth]{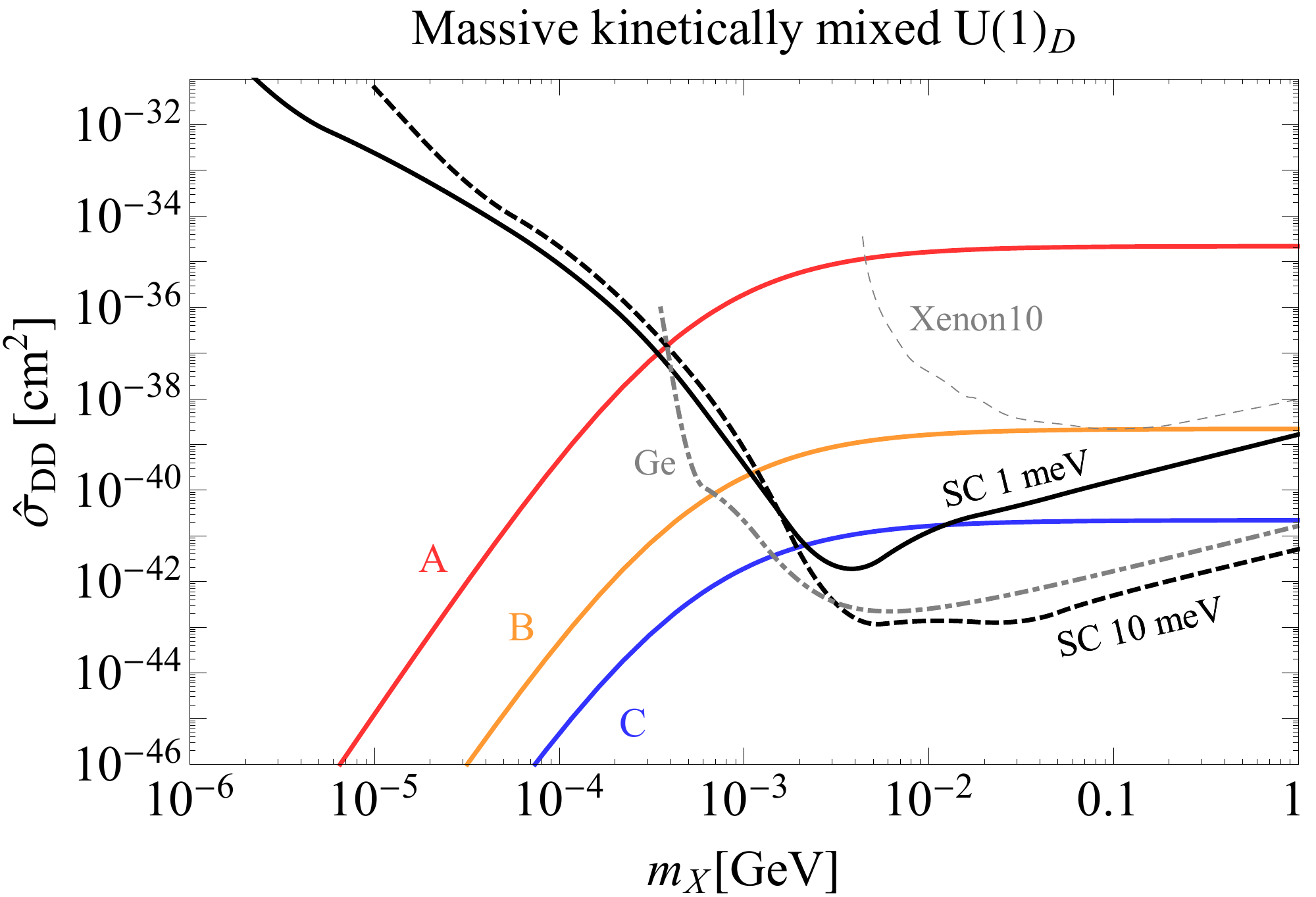}
\caption{ \label{fig:DDcombKIN} {\bf Top:} Upper bounds on the
direct detection cross section, Eq.~\eqref{eq:sigDDhat}, for light
DM scattering off electrons via a kinetically mixed hidden photon,
which obtains its mass via the Stuckelberg mechanism, for a variety
of different mediator masses (solid colored curves). Constraints
include stellar cooling \cite{An:2013yfc},
CMB~\cite{Jaeckel:2010ni}, CROWS~\cite{Betz:2013dza,Graham:2014sha},
measurements of Coulomb's law~\cite{Jaeckel:2010ni}, decoupling at
recombination \cite{Silk:1967kq,McDermott:2010pa} and
self-interactions~\cite{Tulin:2012wi}. {\bf Bottom:} Direct
detection cross section between light DM and electrons, for several
benchmarks of heavy mediators (same as in Fig.~\ref{fig:DDlightS}).
These are {\bf A:} $m_\phi=1$~MeV, $g_e=10^{-5}e$, $\alpha_X=0.1$;
{\bf B:} $m_\phi=10$~MeV, $g_e=10^{-5}e$, $\alpha_X=0.1$; and {\bf
C:} $m_\phi=100$~MeV, $g_e=10^{-4}e$, $\alpha_X=0.1$. These depicted
parameters obey all terrestrial and astrophysical constraints,
though sub-MeV DM interacting with SM through a massive mediator may
be strongly constrained by BBN; see text for details. {\bf In both
panels}, the Xenon10 electron-ionization data
bounds~\cite{Essig:2012yx} are shown in thin dashed gray. The black
solid (dashed)curve depicts the sensitivity reach of the proposed
superconducting aluminum devices, for a detector sensitivity to
recoil energies between 1~meV$-$1~eV (10~meV$-$10~eV), with a
kg$\cdot$year of exposure.  We have included only the solar
neutrino background in our estimate. For comparison,
the gray dot-dashed curve depicts the expected sensitivity utilizing
electron ionization in a germanium target as obtained in
Ref.~\cite{Essig:2015cda}. }
\end{center}
\end{figure*}

If the charge of the dark Higgs is substantially smaller than that
of the DM, or if the hidden photon obtains its mass through the
Stuckelberg mechanism, stellar constraints on the mediator are lifted as $m_{A'}\to
0$ as discussed below Eq.~(\ref{eq:stellarU1}). Considering
the strongest amongst all constraints, we plot in the top panel of
Fig.~\ref{fig:DDcombKIN} the upper bound on $\hat \sigma_{\rm DD}$
in this case for several sample mediator masses, shown in the solid colored curves.
(We note that for very light hidden photon mediators, stellar emission of DM beneath 100~keV severely suppresses the allowed cross section (see Fig.~\ref{fig:mili} below); for this reason, only $m_{\rm DM}\gsim 100$~keV is shown here. In this region, constraints from SN emission of the DM can be released via trapping effects, and so do not control the largest allowed cross section, which we show.)
As is evident, in
contrast to the dark Higgs case, large direct detection cross
sections are possible for the Stuckelberg case.

{\bf Heavy mediator.}
In the bottom panel of Fig.~\ref{fig:DDcombKIN} we plot
several benchmark points, labeled {\bf A-C}, shown in solid colored curves. These theory benchmark curves are the same as those
presented in the bottom panel of Fig.~\ref{fig:DDlightS}, modified to $\hat\sigma_{\rm DD}$ here.
As was the case for the massive scalar/non-kinetically mixed vector mediator, if DM is
lighter than approximately 100 keV, then it must either be a real
scalar or be thermally unpopulated at BBN in order to satisfy
constraints. Alternatively, its couplings and/or mass can vary
within the thermal history of the universe. (Similar statements hold for the very light mediator as well, shown in the cyan curve in the top panel of Fig.~\ref{fig:DDcombKIN}, for very light dark matter masses.)

{\bf Reach.}
The $95\%$ expected sensitivity reach for a kg$\cdot$year of our proposed
superconducting aluminum experiment is shown in the thick black
curves in Figs.~\ref{fig:DDhiggs} and~\ref{fig:DDcombKIN}. The dashed [solid] curves show
the sensitivity when operating with a 10 meV to 10 eV [1 meV to 1
eV] dynamical range. The reach of the superconducting devices for
both light and heavy mediators differs from that in
Fig.~\ref{fig:DDlightS} because the plasma effects are important.
(Note that the effectively massive behavior of the mediator
in-medium results in a better reach for the 10~meV$-$10~eV dynamical
range compared to the 1~meV$-$1~eV range; this is because the rate is
now peaked at higher energy deposits.) As is evident from
Figs.~\ref{fig:DDhiggs} and~\ref{fig:DDcombKIN}, while the
superconducting metal target is not appropriate for detecting some
classes of kinetically mixed light hidden photon models, it is
capable of probing others. When the kinetically mixed photon obtains its mass via a dark Higgs mechanism, superconductors are not ideal DM detectors unless the dark Higgs charge is substantially suppressed compared to that of the DM. In contrast, a kinetically mixed photon with Stuckelberg mass could allow for DM detection via superconductors. The in-medium effects of the metal hurt the low-DM
mass reach due to the large plasma mass of the photon, and the reach
of a semi-conductor target such as germanium or silicon is comparable to the
superconducting devices for masses above a few hundred keV. Below
that, where semi-conductors lose sensitivity, our detectors could be
sensitive to DM masses above 100~keV that scatter by a kinetically mixed $U(1)_D$ with very small Stuckelberg mass.

\subsection{Milli-charged dark matter}\label{ssec:mili}

We now analyze the reach of our method into the parameter space of milli-charged DM particles $X$ with electromagnetic charge $Q$. The `mediator' between the DM and the visible sector is simply the photon, where the DM couples to the electromagnetic current with strength $Qe$. In our notation, this means $g_e = e$ and $g_X=Qe$.

\begin{figure}[t!]
\begin{center}
\includegraphics[width=0.7\textwidth]{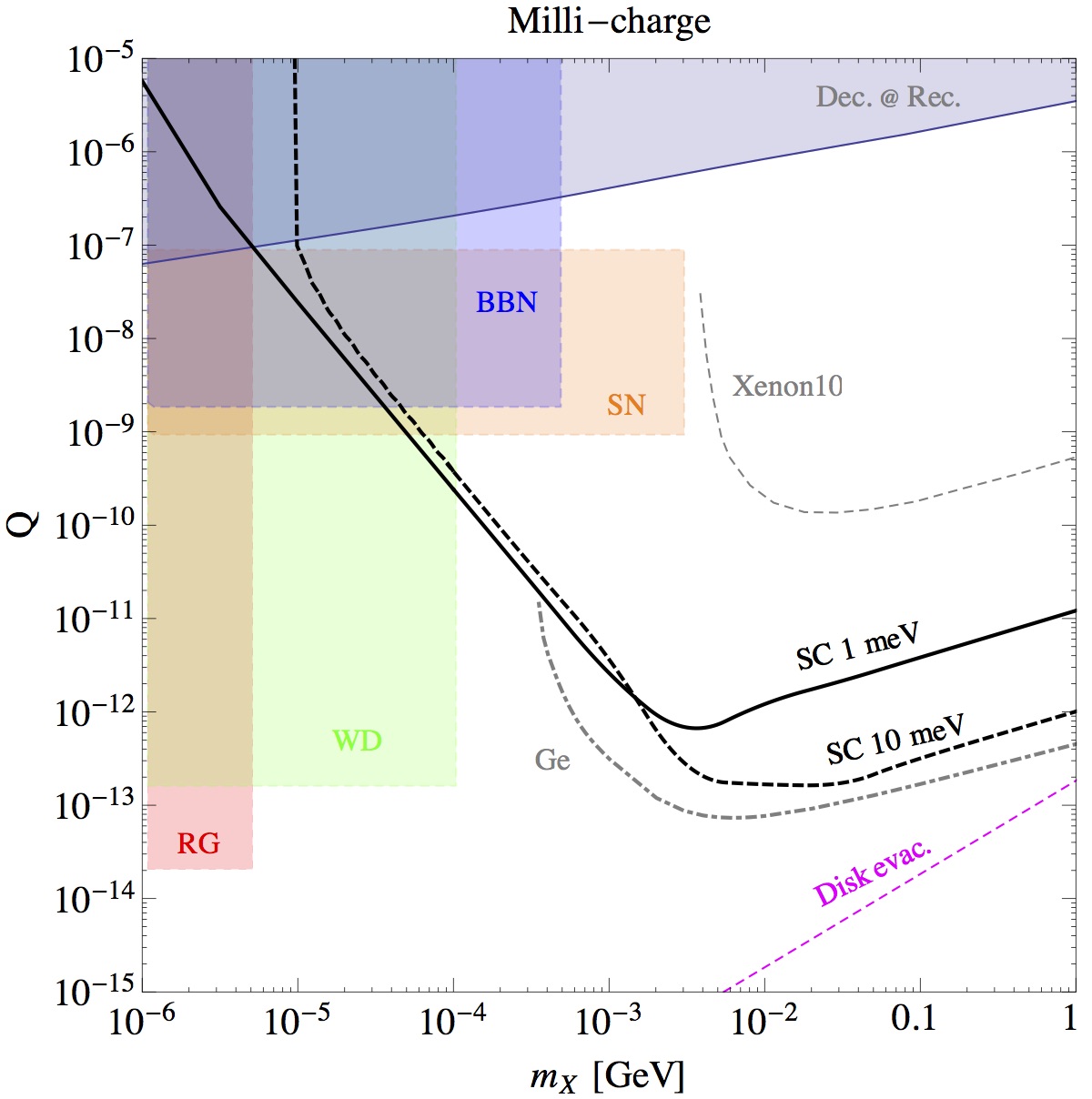}
\caption{Constraints and reach for milli-charged DM: stellar
emission bounds from red giants~(red), white dwarfs~(green) and
supernovae~(orange)~\cite{Davidson:2000hf};
BBN~(blue)~\cite{Davidson:2000hf}; decoupling at
recombination~(gray)~\cite{McDermott:2010pa}; Xenon10~(gray thin
dashed curve)~\cite{Essig:2012yx}; evacuation from the disk~(dashed
magenta curve)~\cite{McDermott:2010pa}; projected reach using a
germanium target~(thick gray dot-dashed curve)~\cite{Essig:2015cda};
expected reach with a superconducting aluminum device with a
sensitivity to recoil energies between 10~meV - 10~eV~(black thick
dashed curve) and 1~meV - 1~eV~(black thick solid curve).  We have included only the solar
neutrino background in our estimate.  }\label{fig:mili}
\end{center}
\end{figure}

The potential reach of the superconductoing devices we propose can
easily be translated into the $Q-m_X$ plane. Constraints on the
milli-charge $Q$ as a function of the DM mass $m_X$ have been worked
out extensively in the literature. Stellar cooling from red giants
(RG), white dwarfs (WD) and supernovae (SN) as well as big bang
nucleosynthasis (BBN) are worked out in Ref.~\cite{Davidson:2000hf}.
In addition, Ref.~\cite{McDermott:2010pa} considers the requirement
of DM be decoupling from the plasma at the time of recombination.
The possibility of charged DM being evacuated from the disk was also
considered in Ref.~\cite{McDermott:2010pa}, though the argument
leading to the constraint is not bullet-proof. In the mass range of
interest, Xenon10 constraints exist as well~\cite{Essig:2011nj,
Essig:2012yx}. These existing constraints are depicted in
Fig.~\ref{fig:mili}, along with the potential reach of our proposed
method. For completeness, we show the projected reach using a semi-conductor germanium
target as well~\cite{Essig:2015cda}; silicon performs similarly.
For masses above a few hundred keV, a germanium/silicon experiment can
outperform superconductors, while for lower masses, where
semi-conducting targets loose sensitivity, the large in-medium
effects of a photon in a metal suppress the reach of
superconductors into the milli-charged DM parameter space. A viable
region can be probed, though the region can be broadened if stellar
and/or BBN constraints are lifted. For milli-charged DM, we learn
that an insulating target would perform better.

\subsection{Dipole operator dark matter}\label{ssec:dipole}

It is possible that DM particles directly couple to photons through
a magnetic or electric dipole operator, which for Dirac fermion DM $X$ takes the form
\begin{eqnarray}
 \label{eq:dipoleL}
 {\cal L}_{\rm dipole}=\frac{1}{2}\bar{X}\sigma_{\mu\nu}(\mu +d\;\gamma_5)X\;F^{\mu\nu}\,.
\end{eqnarray}
Both operators above induce similar physics effects in stellar
cooling processes, and so can be described via an effective dipole moment,
\begin{eqnarray}
 \label{eq:effectDipole}
\mu_{\rm DM}^2=\mu^2+d^2\,.
\end{eqnarray}

For $m_X\lsim {\cal O}({\rm keV})$, the induced DM emission
processes from the Red Giant (RG) branch in globular clusters place a strong bound of~\cite{Haft:1993jt}
\begin{eqnarray}\label{DipoleRG}
\mu_{\rm DM}\lsim 3 \times 10^{-12}\mu_B \quad [{\rm RG}]\,,
\end{eqnarray}
where $\mu_B=\frac{e}{2m_e}\simeq 300$~GeV$^{-1}$ is the Bohr magneton.
For heavier DM, emission from White Dwarfs (WD) and supernova are relevant.
WD cooling restricts~\cite{Bertolami:2014noa}
\begin{eqnarray}\label{DipoleWD}
\mu_{\rm DM}\lsim 5 \times 10^{-12}\mu_B\quad [{\rm WD}]\,,
\end{eqnarray}
which is comparable to the RG constraint, but applicable for a wider range of masses, $m_X\lsim {\cal O}({\rm MeV})$.
For heavier DM with masses up to ${\cal O}(100\; {\rm MeV})$, supernovae provide the best constraints, and the allowed range is ~\cite{Kadota:2014mea}
\begin{eqnarray}\label{DipoleSN}
\mu_{\rm DM}\gsim 2 \times 10^{-11}\mu_B\, \quad {\rm or} \quad \mu_{\rm DM}\;\lsim \; 2 \times 10^{-12}\mu_B\quad [{\rm SN}]\,.
\end{eqnarray}
Above, the upper limit on $\mu_{\rm DM}$ comes from emission
considerations, while the lower limit arises from trapping effects,
which kick in as the coupling between DM and ordinary matter
increases and release the constraints. We note that while the analysis of
Ref.~\cite{Kadota:2014mea} does not include
Pauli blocking effects, the above bound converts into an
effective suppression scale of the dimension-five dipole operator of
order $\sim 10^9$~GeV, which is comparable to the constraint on the
suppression scale of the axion-photon coupling from supernova cooling
considerations~\cite{Raffelt:1996wa}. We thus believe the order of
magnitude of the constraint is valid.

The DM-electron scattering cross sections from Eq.~\eqref{eq:dipoleL} scale as~\cite{Sigurdson:2004zp}
\begin{eqnarray}
 \label{eq:DipoleXsec}
\frac{d\sigma_{\rm E-dipole}}{d\Omega}\propto \frac{d^2}{v_X^2}\,,\quad \quad
\frac{d\sigma_{\rm M-dipole}}{d\Omega}\propto \mu^2\,,
\end{eqnarray}
where the electric dipole scattering is enhanced by $1/v_X^2$, and no low-velocity
enhancement is present for scattering through a magnetic
dipole. This is in contrast to the scalar/vector mediator cases and milli-charged DM,
where the scattering cross section enjoys a
low-velocity enhancement of $1/v_X^4$.
Comparing the milli-charged and electric dipole cases, we have roughly
\begin{eqnarray}
 \label{eq:Comparison}
\frac{\sigma_{\rm milli}}{\sigma_{\rm E-dipole}}&\sim&
\frac{Q^2/v^4}{\mu_{eX}^2 d^2/v^2}\,,
\end{eqnarray}
where $Q$ is the milli-charge of the DM and $\mu_{eX}$ is the reduced
DM-electron mass as usual. Taking $v_X\sim 10^{-3}$, we have
\begin{eqnarray}
 \label{eq:Comparison}
\frac{\sigma_{\rm milli}}{\sigma_{\rm E-dipole}}&\sim&
\left(\frac{Q}{10^{-17}}\right)^2\left(\frac{10 \; {\rm
keV}}{\mu_{eX}}\right)^2\left(\frac{10^{-9}\; {\rm
GeV}^{-1}}{d}\right)^2\,.
\end{eqnarray}
Thus we see that for DM of 10~keV mass, an experiment which is
sensitive to DM with milli-charge as small as $10^{-17}$ can be useful
in probing the unconstrained parameter space in the electric dipole
operator DM scenario. The reach for magnetic dipole operator is
worse since there is no velocity enhancement. Comparing to
Fig.~\ref{fig:mili}, we learn that the proposed superconducting
detectors will not be sensitive to dipole DM.

\section{Conclusions}\label{sec:conc}

We have explored in detail a proposal for detecting DM with
Fermi-degenerate materials, focused on the case of a superconducting
metal target.  We
computed the scattering rate of DM off of the electrons, factoring in the suppression due to
Pauli blocking effects.  We considered cosmological and astrophysical constraints
from DM self-interactions, kinetic decoupling in the early universe as well as
stellar emission, together with terrestrial constraints, such as $(g-2)$ of the
electron and
beam dump experiments.  These constraints were then applied
to a variety of models, such as a simplified model of a scalar or
vector mediator, a kinetically mixed $U(1)_D$ and milli-charged DM.
We have shown that viable regions of parameter space exist, consistent with various
cosmological, astrophysical and terrestrial constraints, which are detectable with our proposed experiment.  A broader range of model
parameter space becomes available if stellar and/or $N_{\rm eff}$ constraints
on light degrees of freedom are lifted; we leave the exploration of such models for future work.
We also computed in-medium effects for the kinetically mixed dark
$U(1)_D$, and found that the plasma mass of a photon in a metal substantially reduces the reach of superconducting detectors for this class of models.

There are several further directions that we are pursuing.  First,
since the reach in a metal is reduced for a kinetically mixed dark
$U(1)_D$, other types of target materials should be examined which feature
small or zero energy gap and simultaneously also small in-medium photon mass; graphene is
one possibility. Second, in this paper our attention was restricted
to $t$-channel scattering of DM off the target.  We plan also to examine, however,
the absorption of very low mass states via the excellent energy
resolution of the experiment.  Third, while we focused here on targets with a substantial
initial state velocity, in order to find configurations where the entire kinetic energy of the DM can be extracted, we also note that ${\cal O}({\rm meV})$ energy deposits on a light nucleus could also be utilized for probing light DM.  We are currently pursuing helium targets as well.

Over the last decades, the main focus of the DM community has been directed towards
axions and the weak scale as the source of DM.  As the pursuit
in the search for DM expands, it is important to consider as
broadly as possible what types of DM experiments can be built and what
types of models, consistent with all known constraints, could be
detectable.  Exploiting superconducting targets is an important step along this path.

\section*{Acknowledgments}
We thank Ehud Altman, Haipeng An, John Clarke, Snir Gazit, Roni
Ilan, Eric Kuflik, Tongyan Lin, Dan McKinsey, Dave Moore, Joel Moore, Maxim Pospelov, Zohar
Ringel and Kai Sun for very useful discussions. The work of YH is
supported by the U.S. National Science Foundation under Grant No.
PHY-1002399. YH is an Awardee of the Weizmann Institute of Science
-- National Postdoctoral Award Program for Advancing Women in
Science. YH thanks the Aspen Center for Physics where part of this
work was done, supported by NSF grant PHY-1066293. YZ is supported
by DE-SC0007859. KZ is supported by the DoE under contract
DE-AC02-05CH11231.

\begin{appendix}

\section{Relation between conductivity, index of refraction and $\Pi_{L,T}$ in medium}\label{sec:app}

We start with the relation~\cite{Schreiffer}
\beq J_\mu(q) = - R_{\mu \nu}(q)
A_\nu(q)\,, \label{CurrentVecPot}
\eeq
where
\beq
{\rm Re}~\Pi_{\mu\nu}({\bf q},q_0) = {\rm Re}~R_{\mu\nu}({\bf q},q_0)\,, \\
{\rm Im}~\Pi_{\mu\nu}({\bf q},q_0) = {\rm sgn}(q_0) {\rm
Im}~R_{\mu\nu}({\bf q},q_0)\,. \eeq We now use Maxwell's equations to
write this in terms of the conductivity.  We know that longitudinal
conductivity is defined by $\vec{J} = \sigma_L \vec{E}$.  We also
know that current conservation dictates $\partial_\mu J^\mu = 0$
implying $\omega J^0 = q_i J_i$ (we will use an Einstein summation convention and roman letters to denote spatial indices in this Appendix).  We can thus write, using
Eq.~\eqref{CurrentVecPot} and the fact that $R_{0i} A_i = 0$ in Coulomb gauge,
\beq
\frac{\sigma_L}{\omega}q_i E_i = -R_{00}A_0\,.
\eeq
We also have from Maxwell's equations in Coulomb gauge $\vec\nabla \cdot
\vec E = -\nabla^2 \phi$. Identifying $\phi = A_0$, this allows us
to write
\beq \frac{i
{\bf q}^2\sigma_L}{\omega} = -R_{00}\,. \eeq
Using $\Pi_{00} =
{\bf q}^2/q^2 \Pi_L$ [from Eqs.~\eqref{PolarizationTensor} and~\eqref{eq:eps_tensor}] and Eq.~\eqref{SigmaL}, we recover the longitudinal part of Eq.~(\ref{PiLPiT}).

Now we turn to the transverse component, for which \beq J_i = -
R_{ij}(q) A_j(q)\,. \eeq
We use Maxwell's equation $\vec{E} = -
\frac{\partial \vec{A}}{\partial t } - \vec{\nabla} \phi = i \omega
\vec{A} - i \vec{q} A_0$.  This allows us to
write
\begin{eqnarray}
J_i = - \frac{1}{i \omega} R_{ij}[E_j + i q_j A_0]\,.
\end{eqnarray}
Then from the first Maxwell equation $\vec\nabla \cdot \vec E =
-\nabla^2 \phi$ we can further write
\beq J_i = -\frac{1}{i \omega}
R_{ij}[\delta_{j\ell} - \frac{q_j q_\ell}{{\bf q}^2}] E_\ell\,.
\eeq
Using $2 \sigma_T = P_{T ij} \sigma_{ij}$, we learn that the transverse conductivity is
\beq
\sigma_T =-\frac{1}{2 i \omega} [\delta_{j\ell} - \frac{q_j q_\ell}{{\bf q}^2}] R_{j \ell}.
\eeq
Identifying $\Pi_T = [\delta_{j\ell} - \frac{q_j q_\ell}{{\bf q}^2}] R_{j \ell}$, we recover the transverse part of Eq.~\eqref{PiLPiT}.

\end{appendix}


\end{document}